\documentclass[fleqn,usenatbib]{mnras}

\usepackage{newtxtext,newtxmath}

\usepackage[T1]{fontenc}
\usepackage{amsfonts}
\usepackage{graphicx}
\usepackage{subcaption}
\usepackage{hyperref}

\DeclareRobustCommand{\VAN}[3]{#2}
\let\VANthebibliography\thebibliography
\def\thebibliography{\DeclareRobustCommand{\VAN}[3]{##3}\VANthebibliography}

\title[Impact of Stirring on Discs: A Study of HD~16743]{Testing the Impact of Planet-stirring, Self-stirring, and Mixed-stirring on Debris Disc Architecture: A Case Study of HD~16743}

\author[Marshall et al.]{
Jonathan P. Marshall,$^{1}$
Marco A. Mu\~noz-Guti\'errez,$^{2}$\thanks{Corresponding author, e-mail: marco.munoz@uda.cl (MAM)}
Antranik A. Sefilian$^{3}$
and A. Peimbert$^{4}$
\\
$^{1}$Institute of Astronomy and Astrophysics, Academia Sinica, 11F of AS/NTU
Astronomy-Mathematics Building, No.1, Sec. 4, Roosevelt Rd, Taipei 106319, Taiwan\\
$^{2}$Instituto de Astronom\'ia y Ciencias Planetarias, Universidad de Atacama, Copayapu 485, Copiap\'o, Chile\\
$^{3}$Department of Astronomy and Steward Observatory, University of Arizona, 933 N Cherry Ave, Tucson, AZ 85721, USA
\\
$^{4}$Instituto de Astronom\'ia, Universidad Nacional Aut\'onoma de M\'exico, Apdo. postal 70-264, Ciudad Universitaria, M\'exico
}

\date{Accepted XXX. Received YYY; in original form ZZZ}

\pubyear{2025}

\begin{document}
\label{firstpage}
\pagerange{\pageref{firstpage}--\pageref{lastpage}}
\maketitle

\begin{abstract}
Dynamical interactions between planets and debris discs can excite the orbits of embedded planetesimals to such a degree that a collisional cascade is triggered, generating detectable amounts of dust. Millimetre wavelength observations are sensitive to emission from large and cold dust grains, which are unperturbed by radiation forces and act as a proxy for the location of the planetesimals. The influence of unseen planetary companions on debris discs can be inferred with high-resolution imaging observations at millimetre wavelengths, tracing the radial and vertical structure of these belts. Here we present a set of $N$-body simulations modelling ALMA observations of the HD~16743 debris disc. We consider a range of relative contributions from either a single planetary companion and/or a set of embedded massive planetesimals to reproduce the disc's observed radial and vertical structure. We compare our dynamical results for the limiting cases of planet-stirring and self-stirring, finding them to be consistent with theoretical expectations for each scenario. For the case of HD~16743, we find that a set of massive planetesimals on mildly eccentric orbits, confined to a relatively narrow range of semimajor axes (compared to the observed belt width), offers the best results to reproduce the vertical and radial extent of the observed emission.
Our findings constrain the total planet–disk system mass. A combined giant and dwarf planet mass of $\geq~27~M_{\oplus}$ can reproduce the observed architecture, with the equipartition scenario requiring only half the disc mass of the self-stirring scenario.
\end{abstract}




\section{Introduction}

HD~16743 is a relatively young F-type star at a distance of $57.807\pm0.060$~pc \citep{Gaia18} with an estimated age around 60~$\pm$~20~Myr, but likely below 500~Myr \citep{Desidera15,Marshall23}. The star hosts a relatively bright debris disc ($L_{\rm d}/L_{\star} \simeq 10^{-4}$), which was detected by \textit{Spitzer} \citep{Moor11} and marginally resolved by \textit{Herschel} \citep{Moor15,Marshall21}. The debris disc emission is consistent with a two-component structure based on the spectral energy distribution modelling, a brighter and more massive analogue to the Solar system's Asteroid and Kuiper belts \citep{Horner20}. From the far-infrared \textit{Herschel} observations, estimates of the mass and semi-major axis of any planetary companion interacting with the disc from \citet{Pearce22} encompass a 4~$\pm$~10~$M_{\rm Jup}$ planet interior to 80~au (to stir the disc), or a 20~$M_{\oplus}$ planet at 110~au (to sculpt the inner edge of the disc). Most recently, the disc was spatially resolved at millimetre wavelengths by ALMA in Band 6 ($\simeq 1.3~$mm), from which a radius of 158~au and a width of 34~au (standard deviation; FWHM 79~au) were determined under the assumption of a Gaussian belt model for the architecture \citep{Marshall23}. The disc was further found to be highly inclined ($i = 87 \fdg 3^{+1.9}_{-2.5}$) with a relatively extended vertical aspect ratio $h = z/R = 0.13$. This is potentially indicative of stirring of the planetesimal belt by a planetary companion or planetesimals within the debris disc \citep{Marshall23}.

A debris disc is stirred if its particles have considerable relative velocities that allow collisions among them to be erosive or even disruptive \citep{Mustill09,2010Krivov}. This is attained with sufficiently large values of the particles' eccentricities and inclinations, but such values have to be the result of a dynamical excitation process, assuming debris discs are unstirred following the dissipation of the gas in the progenitor proto-planetary disc. In general, two methods of stirring have been considered, first, secular perturbations produced by an internal or external giant planet on an eccentric and/or an inclined orbit, which in turn drives planetesimals onto eccentric and inclined orbits \citep[e.g.][]{Mustill09,Pearce22, Farhat23}, and second, a direct self-stirring process induced by the largest planetesimals (dwarf planets) present in the debris disc, which increases eccentricities and inclinations by direct scattering perturbations over the less massive members of the disc \citep[e.g.][]{Kenyon08,Krivov18}. Both scenarios, though successful in modelling certain systems, run into problems when a more general picture is invoked. In particular, the secular stirring scenario requires an already excited large perturber, while the self-stirring option sometimes requires an unreasonably large mass of the disc itself to be effective \citep{WyattKrivov21}. For instance, the disc mass required for HD~16743's disc to be self-stirred has been estimated in a range $\simeq18~M_{\oplus}$ \citep{Marshall23} to 3,000~$\pm$~2,000~$M_{\oplus}$ \citep{Pearce22}. This large span stems from the different belt architectures and assumptions of the self-stirring model, and the inherent uncertainties on the planetesimals responsible for the disc stirring. For example, \citet{Pearce22} adopted the disc architecture from marginally resolved \textit{Herschel} observations, assuming a young age based on Argus association membership \citep[45~$\pm$~5~Myr;][]{Zuckerman19}, and used a stirring model in which the massive planetesimals are small, on circular orbits, and non-interacting \citep[following][]{Krivov18}. These factors would elevate the disc mass required to stir the disc, whereas \citet{Marshall23} following the stirring model of \citet{2012Pan}, consider only kinetic energy arguments to infer their much lower mass estimate for the disc, or rather its perturber(s). However, their model neglects the time required to excite the disc across its width. Moreover, recent findings suggest that if the disc is sufficiently massive, the collective gravity of the planetesimals can suppress -- or at least mitigate -- the secular perturbations typically induced by an eccentric and/or misaligned planet \citep[][]{Sefilian2024, Sefilian2025}.

Some variations of the above mechanisms have recently been considered to alleviate the limitations of these traditional scenarios \citep[e.g.][]{Munoz15,Costa23}. In particular, \cite{Munoz23} have considered in detail the combination of giant and dwarf planetary perturbations over a dynamically cold debris disc, finding that, in general, a simple combination is more efficient in stirring a debris disc than either ingredient acting in isolation. This scenario is referred to as ``mixed-stirring'' to distinguish it from the secular and self-stirring mechanisms described above \citep{Munoz23}. 

In this work, we use HD~16743 as a test bed for the mixed-stirring scenario. We adopt this system because its measured vertical aspect ratio lies in the mid-range of the ensemble of systems with measured values \citep[$\sim$0.02--0.20;][]{Terrill23}, which, combined with its relatively young age and broad width, should lead to less ambiguity regarding the disc's stirring mechanism. Our main goal is to determine whether the level of dynamical excitation in the disc implied by the ALMA observations can be reproduced by a relatively low-mass planet acting in combination with dwarf planets embedded in the disc. We will show that a single initially dynamically cold planet 
(i.e., on a near-coplanar and circular orbit) with a mass consistent with the observed constraints for the HD~16743 system is ineffective in stirring this disc across its full width within the age of the system. We will also show that the degree of dynamical excitation produced by a self-stirring disc can equivalently be achieved by a combination of an initially dynamically cold planetary companion and a less massive debris disc working in concert. 

The remainder of the paper is laid out as follows. In Section \ref{sec:methods}, we describe the dynamical models and synthetic observations used in this analysis. Then, we present the results of our $N$-body simulations in Section \ref{sec:results_dyn} and of the synthetic observations in Section \ref{sec:results_obs}. We discuss the implications of these results in Section \ref{sec:discussion}, before finally presenting our conclusions in Section \ref{sec:conclusions}.

\section{Methods and Simulations}
\label{sec:methods}

To explore the stirring mechanism(s) that shape the debris disc observed around HD~16743, we conducted a series of numerical simulations using the $N$-body code {\sc rebound} \citep{rebound}. The primary objective of our simulations is to assess the applicability of the mixed-stirring scenario proposed by \citet{Munoz15, Munoz23} to a real system. In this scenario, dwarf planets, which are at the upper end of the size distribution of debris discs and contain most of the disc's mass, act as perturbing agents alongside an interior giant planet. Together, they drive the long-term orbital evolution of test or dust particles within an initially cold debris disc.

\subsection{Orbital configuration of the simulated systems}

\begin{figure*}
    \centering
    \includegraphics[width=0.8\textwidth]{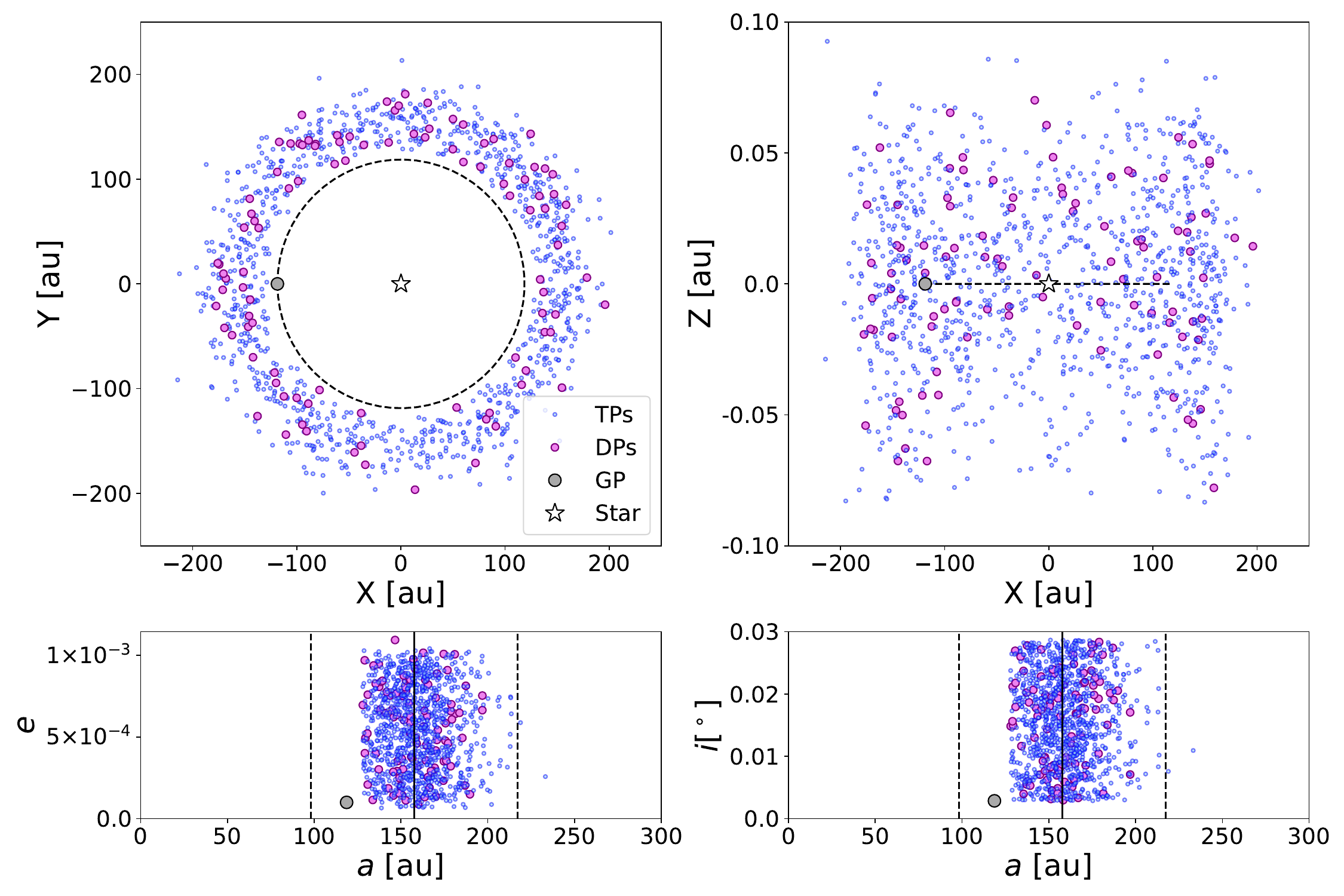}
    \caption{Initial conditions of a representative simulation from the model grid: $M_{\rm tot} = 27~M_{\oplus}$, with 50\% of the mass in DPs, and $\sigma = \mathrm{FWHM}/4$. \textit{Top}: Distribution of instantaneous positions of test particles (TP; blue), dwarf planets (DP; purple), giant planet (GP; black), and the star (white) in $X$--$Y$ and $X$--$Z$ planes. For reference, the planet's orbit is denoted by the dashed line. Note the change in scale between the vertical axes of the top left and top right panels. \textit{Bottom}: 
    Distribution of eccentricities $e$ and inclinations $i$ as a function of semimajor axis $a$ for all bodies involved (TPs, DPs, and the GP).
    The vertical solid and dashed lines denote the peak and extent of the planetesimal belt ($\mu_{0}~\pm~3\sigma_{\rm disc}$, where $\sigma = \mathrm{FWHM}/4$). TPs and DPs located within 3 Hill radii of the planet were removed and their orbital elements redrawn from the distributions, leading to the truncation of the disc's inner edge.}
    \label{fig:fiducial_sim}
\end{figure*}

Throughout this work we consider systems composed of a central star with mass  $M_c=$ 1.537~M$_\odot$ \citep[adopted from][]{Marshall23}, and a giant planet located at several tens of au, with its precise location depending on the width of the disc (more on this below). The planet is assumed to be initially on a nearly circular, coplanar orbit interior to a debris disc (namely, with $e_{\rm GP}= 2 i_{\rm GP} = 10^{-4}$).\footnote{Note that when the giant planet is the only or the overwhelmingly more massive body, its angular momentum defines the reference plane of the system as a whole.} We show an example of the initial conditions in Figure \ref{fig:fiducial_sim}. 
The remainder of this subsection outlines the reasoning behind the adopted initial configuration.

To model HD~16743's debris disc, we use the results of \cite{Marshall23}; they found that the disc's centre, $\mu_0$, is located at 157.8 au from the star, with a full width at half maximum (FWHM) of 79.4 au (equivalent to standard deviation $\sigma$ = 33.7 au). In this light, we considered an initially dynamically cold debris disc formed by 100 massive planetesimals (a.k.a. dwarf planets, hereafter DPs) and 1000 test particles (hereafter TPs). The initial semimajor axes of massive and massless particles in the disc are chosen from a random log-normal distribution, such that $\log(a_{\rm DPs})$, and equivalently $\log(a_{\rm TPs})$, follow a normal distribution with mean $\mu_0$ and standard deviation $\sigma$. We consider a range of disc widths based on a fraction of the measured FWHM between 1/2 and 1/8th of the measured value.
The semi-major axes of TPs and DPs are thus drawn from a log-normal distribution with a peak 
\begin{equation}
    \log \mu = \log \frac{\mu_{0}^{2}}{\sqrt{(\mu_{0}^{2} + \sigma^{2})}}
\end{equation}
and width 
\begin{equation}
    \log \sigma = \sqrt{ \log \left( 1 + \frac{\mathrm{FWHM}^{2}}{\mu_{0}^{2}} \right) }
\end{equation}
where $\mu_{0}$ is 157.8~au, and $\sigma$ is the scaled width of the disc FWHM/$n$ where FWHM = 79.4~au and $n = {2,4,8}$.
To avoid a random spread of particles toward the inner region of the system, we defined an inner edge for our initially generated discs based on the value of $\sigma$, which depends on $n$. This inner edge is placed at 39.9, 19.8, or 9.9 au from the centre of the disc for the $n={2,4,8}$ models, respectively. Particles located interior to this limit (within 3 Hill radii of the planet) have their semi-major axis redrawn from the same distribution such that the overall distribution of the belt is preserved without bias.

We consider a set of discs that are increasingly narrower than the observed distribution (for larger values of $n$) at the start of the simulations. This is because bright lobes at the edges of observed debris discs are expected to be produced by narrow belts instead of broad discs, and this morphology is what is observed in the ALMA Band 6 image of HD~16743 \citep[e.g. Figure 1 in][]{Marshall23}. 
Relatedly, the observed radial width of the disc results from the superposition of eccentric orbits due to dynamical stirring, meaning the underlying distribution of semimajor axes should be narrower than the measured radial extent \citep{Rafikov2023} -- although the exact difference is difficult to determine.

The initial eccentricities and inclinations of both TPs and DPs are drawn randomly between $e_{\rm min}=10^{-4}$ and $e_{\rm max}=10^{-3}$, and between $i_{\rm min}=e_{\rm min}/2$ ($= 5\times10^{-5}$) and $i_{\rm max}=e_{\rm max}/2$ ($= 5\times10^{-4}$), respectively.\footnote{We opt to such small values of eccentricities and inclinations to make sure they are significantly smaller than the critical values required for the disc to be considered stirred; see Section \ref{sec:results_dyn} for further details.} Finally, for both TPs and DPs, the remaining angular elements, namely the argument of pericentre, $\omega$, the longitude of the ascending node, $\Omega$, and the mean anomaly, MA, are randomly drawn between 0 and $2\pi$. 

The initial semimajor axis of the planet depends in each case on the radial width of the disc and the mass of the planet $m_{\rm GP}$. In all cases, the planet is placed 3 Hill radii away from the inner edge of the disc, i.e. $a_{\rm GP} = R_{\rm in} - 3 R_{\rm Hill}$, where 
\begin{equation}
    R_{\rm Hill}   \approx     a_{\rm GP}  \bigg( \frac{m_{\rm GP}}{3 M_c} \bigg)^{1/3}      .
\end{equation}
Depending on the model, this corresponds to a semimajor axis between 90 and 135 au. The planet is initialized with an eccentricity of $e_{\rm GP}=10^{-4}$ and an inclination of $i_{\rm GP}=e_{\rm GP}/2$. The planet's longitude of periastron ($\omega$) and ascending node ($\Omega$) are initially set to $0$, and its mean anomaly (MA) is set to $180\degr$.

\subsection{Physical properties of the planets and the discs}

In \cite{Munoz23}, it was found that the mass ratio between the giant planet and the debris disc is a crucial factor in determining the stirring level achieved by an initially dynamically cold system. Building upon this idea, we consider the total mass of our systems to be given by $M_{\rm tot}=M_{\rm GP} + M_{\rm DD}$, where $M_{\rm GP}$ is the mass of the giant planet and $M_{\rm DD}$ the mass of the disc, such that $M_{\rm DD}=\sum_{i=1}^{100}M_{\rm DP_{\it i}}$, and $M_{\rm DP_{\it i}}$ is the mass of each DP. 

We explored three values for $M_{\rm tot}$, guided by the estimates of \citet{Marshall23} for HD~16743. Using the collisional cascade model of \cite{2012Pan}, they determined that a total mass of 18~M$_{\oplus}$ is sufficient to reproduce the observed aspect ratio of HD~16743's disc. This estimate considered only kinetic energy arguments from collisional production of dust and did not require a particular timescale of action to reproduce the vertical extent of the disc. The inferred mass could plausibly be contained in one of three ways:
as a single giant planet, the outermost planet in a chain 
and located interior to the disc, or as the total sum of $N$ dwarf planets embedded in the disc. Considering this mass, we explore up to a factor of two over this starting value, that is, we simulate systems with $M_{\rm tot}=\{18, 27, 36\}$~M$_\oplus$.  

As mentioned above, the total mass of our systems is distributed between the giant planet and the 100 DPs. Our grid of simulations then considers four cases, namely 0\%, 10\%, 50\%, and 100\% of $M_{\rm tot}$ distributed among the 100 DPs. In other words, we consider the extreme cases in which $M_{\rm tot}=M_{\rm GP}$ (the 0\% mass in DPs case), or $M_{\rm tot}=M_{\rm DD}$ (the 100\% mass in DPs case, which is equivalent to a self-stirring scenario where all the mass is contained in the disc). The intermediate cases, i.e. 10\% and 50\% of the mass carried out by the DPs, correspond to different levels of the mixed-stirring scenario, with which we aim to find a trend for an optimal mass ratio to reproduce the observed properties of the HD~16743 disc, as well as to differentiate the stirring level produced by a giant planet in a planar and circular orbit, from the more widely studied self-stirring case by embedded DPs.

The mass of each DP is assigned randomly by drawing from a differential mass distribution of the form $dn/dm\propto m^{-2}$. Preliminary exploration of the impact of the exponent of the mass distribution by \citet{Munoz17} revealed no strong dependence (with exponents of 1.8, 2.0, and 2.2), so we did not investigate different mass distributions here. Each DP mass within the distribution is appropriately re-scaled such that $M_{\rm DD}$ accounts for the specified percentage of the system being analyzed, while maintaining the slope of the mass distribution of the massive objects. Additionally, we remove the ten most massive objects from our initial distribution, then rescale the disc mass amongst the next 100 most massive bodies. By removing those 10 bodies we avoid DPs with more than 10\% of the disc mass and guaranteeing that no single DP has more than 20\% of the disc mass. This prevents a scenario where a handful of objects would dominate the dynamics of the disc, or a single massive DP could act in a manner analogous to a second planet embedded in the belt.


Under the above considerations, we obtain a simulation grid consisting of 36 different runs.\footnote{We note that an initial grid consisting of 144 simulations, in which the location of the giant planet was moved closer or farther from the disc was also considered, but we found this parameter to be less relevant than the mass ratio between the planet and the disc \citep{Munoz23}. We decided to focus exclusively on the mass ratio, allowing us to reduce the number of simulations and the computational expense of our experiments.} This number of simulations will enable us to explore the effect on the stirring of debris disc particles (both TPs and DPs) resulting from a mass partition between the giant planet and the disc. 

\subsection{Integration parameters}

Our systems were integrated up to 80~Myr by adopting the preferred age of HD~16743 as found in \cite{Marshall23}, i.e., 60$\pm20$~Myr. We used the hybrid symplectic integrator {\sc mercurius} \citep{reboundmercurius}, included in the {\sc rebound} package \citep{rebound}, with an initial time-step of approximately 5 yr (or 0.01 times the orbital period of the giant planet, $\sim500$ yr). A changeover limit between the symplectic and the IAS15 integrators was set to 3 Hill radii from each massive body. This limit allows close encounters with any massive object to be properly resolved using the adaptive time-step integrator IAS15. 
We used the Simulation Archive format \citep{reboundsa} to save simulation data every 80 kyr, for a total of 1001 data outputs for each simulation. 


\section{$N$-body simulations}
\label{sec:results_dyn}

For a disc to be considered stirred, a minimum eccentricity of disc particles has to be achieved such that the relative velocities among particles are high enough to be predominantly erosive or destructive. A calculation of this minimum eccentricity can be found in \citet{Kobayashi2001} \citep[see also][]{Costa23, Sefilian2024}.
Based on the approach of \citet{Sefilian2024}, the critical limit for stirring to occur lies around $e \simeq 0.01$ for cm-sized particles in a disc at $a\sim 100$ au (see their equation 26). The precise value of critical $e$ depends on the composition of the bodies. However, our simulations exceed the above limit within $\sim$ 1 Myr (see e.g. Figures \ref{fig:fiducial_sim} and \ref{fig:rms_e_i_M27_s4}).
In this work, we instead analyse the stirring level of our discs by examining the evolution of the root mean square (RMS) eccentricity (RMS$_e$). We also study the evolution of the RMS inclination (RMS$_i$) since it is related to the maximum disc aspect ratio, a quantity that we intend to relate to the value deduced from the ALMA image of HD~16743 \citep[$h = 0.13~\pm~0.02$;][]{Marshall23}. 

We extracted the main orbital parameters, $a$, $e$, and $i$ of all the objects (GP, DPs, and TPs) in each model system from our simulations, recorded using the {\sc rebound} Simulation Archive \citep{reboundsa}. We then determine RMS$_e$ and RMS$_i$ for the DPs and TPs at each time step to track the dynamical evolution of the belt. In all our simulations, while not shown here, we observe a rapid increase in the values of both RMS$_e$ and RMS$_i$ from their initial values. Over the initial 0.1 Myr of the simulations, they increase by around two orders of magnitude. Following this rapid growth, the rate of evolution slows down, tending towards a slow and steady growth until the end point of the simulations.

\subsection{A fiducial simulation}

\begin{figure}
    \includegraphics[width=0.5\textwidth]{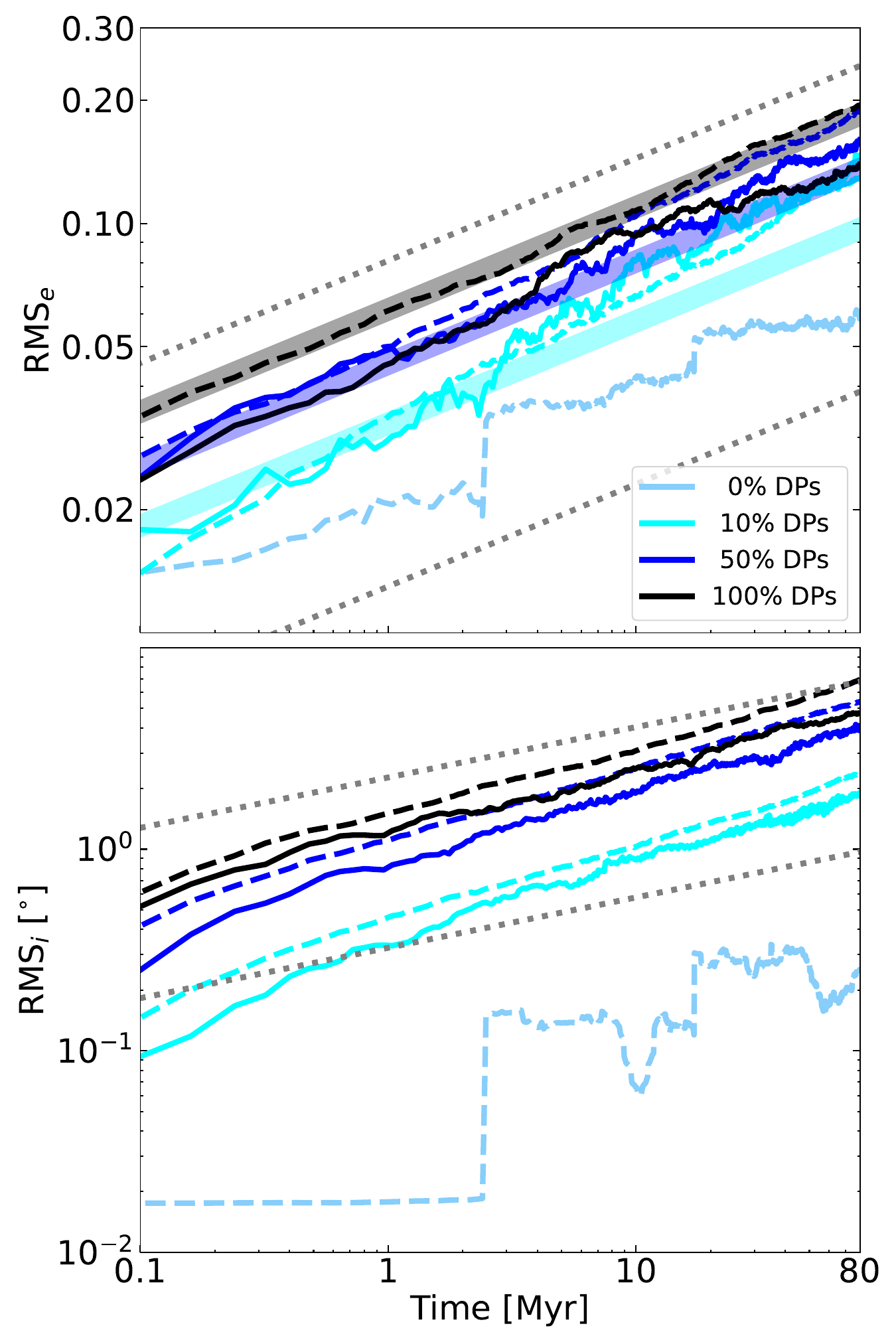}
    \caption{Evolution of RMS$_{e}$ (top) and RMS$_{i}$ (bottom) for the 27~M$_\oplus$ total mass model, with a disc standard deviation $\sigma = \mathrm{FWHM}/4$ (the central model in our grid, see Figures \ref{fig:rms_e} and \ref{fig:rms_i}). The light blue curves denote the 0\% mass in DPs scenario, which barely stirs the disc. The remaining mass fractions are denoted by cyan (10\%), blue (50\%), and black (100\%) lines. Solid lines denote the RMS values for DPs, and dashed lines denote the RMS values for TPs. The shaded regions (top panel) and grey dotted lines (both panels) denote a $t^{1/4}$ dependence increase for the RMS values, based on \citet{Krivov18}.}
    \label{fig:rms_e_i_M27_s4}
\end{figure}

To exemplify the typical evolution of our systems, we choose a fiducial simulation to show in detail. From our grid, we choose the central case with the following parameters: $M_{\rm tot}=27M_\oplus$, $\sigma = \mathrm{FWHM}/4$. In Figure \ref{fig:rms_e_i_M27_s4} we show the evolution of the RMS$_e$ (top panel) and the RMS$_i$ (bottom panel) of our fiducial case, for the different percentages of mass in DPs (or equivalently, the mass of the disc).

From both panels in Figure \ref{fig:rms_e_i_M27_s4} we can see that for the 0\% model, which is comprised of a $27M_\oplus$ giant planet and a massless disc (light blue lines), the value of RMS$_e$ does not increase as much as other simulations, reaching a maximum value of $\sim0.06$.\footnote{Note that for this 0\% case TPs and DPs are equivalent since DPs have no mass and are thus treated as a single population of 1\,100 TPs.} In RMS$_i$, we see large stochastic jumps resulting from individual particles getting close to the perturbing region of the GP (approximately $3~R_{\rm Hill}$), however, disc particles only reach a maximum value of $\sim0\fdg3$. 
In these scenarios with massless DPs, the excitation is predominantly caused by scattering by the planet and interaction with its mean motion resonances \citep{Costa23}.


The above results for the 0\% model are typical among our grid of simulations, and demonstrate the inability of a single giant planet (initially on a circular and coplanar orbit with the disc; an assumption we relax later in Section \ref{subsec:planet_alone_pred}) to stir in any significant amount an initially cold debris disc. In what follows, we will omit showing the evolutionary curves of other 0\% models with different masses of the giant planet. We will focus instead on the evolution of systems with massive discs.


In discs with non-zero mass, one of the first things to note is the dramatic change in the behaviour of both the RMS$_e$ and RMS$_i$ for both particles and DPs. In contrast to the model with a massless disc, we can now see that combining a giant planet and massive perturbers in the disc is much more efficient at exciting the disc particles than the giant planet alone. In fact, from just a modest increment of the mass in the disc (10\% of the total mass considered, shown in cyan lines), the values of RMS$_e$ and RMS$_i$ grow almost monotonically as a power law function of time with an exponent close to 1/4. To exemplify this behaviour, we plotted dotted gray lines showing the function $\propto t^{1/4}$, as a visual reference to compare the behaviour of both quantities, RMS$_e$ and RMS$_i$. We present this function based on the results from \citet{Krivov18} for self-stirring models, which provide an approximate representation of our results, if not in magnitude, at least in their behaviour over time (see Section \ref{subsec:KB18_discussion} for a discussion).

As the fraction of total mass contained in the disc is increased from 10\% (cyan lines) to 50\% (blue lines) and then to 100\% (black lines), the excitation level grows with disc mass. Indeed, we note that the change between the final RMS$_e$ value in a disc of zero mass, to one containing 10\% of the mass of the system, is almost a factor of two (RMS$_e$ increases from approximately 0.06 to 0.11). The same factor of two change in RMS$_e$ is seen between the 10\% and the 50\% models (RMS$_e$ from $\sim$0.11 to $\sim$0.2), while there is no similar increment in RMS$_e$ between the 50\% and the 100\% models. This represents a case of diminishing returns between the increase in mass fraction represented by planetesimals and the level of stirring achieved within the disc.

The behaviour described above does not hold for the evolution of RMS$_i$. First, the change in final RMS$_i$ is of almost an order of magnitude between the massless disc case and the 10\% case models (from approximately 0\fdg3 to 2\fdg5). Also, the maximum RMS$_i$ values reached by TPs continue to rise as the disc mass increases, going from 2\fdg5 to 5\fdg5 to 7$\degr$ for the 10\%, 50\%, and 100\% models, respectively. Thus, for inclinations, no apparent saturation level is reached in our simulations, unlike the case for eccentricities in the same models.

Figure \ref{fig:rms_e_i_M27_s4} also shows that the excitation levels of DPs, both in eccentricity and inclination, consistently lag behind those of TPs. This is expected, as it is more difficult to stir massive particles compared to massless ones. Additionally, there is a noticeable difference in the final RMS$_e$ values achieved by DPs. The 100\% model yields a lower value ($\approx 0.15$) compared to the 50\% model ($\approx 0.2$). In terms of inclinations, there is no apparent saturation limit for DPs; however, their final RMS$_i$ values remain consistently lower than those of TPs.

For comparison, in Figure \ref{fig:rms_e_i_M27_s4} we also show theoretical expectations for the evolution of the RMS$_e$ in the self-stirring scenario from \citet{Ida93,Krivov18}. To evaluate the evolution of RMS$_e$ we apply equation (9) from \cite{Krivov18} to the parameters of our simulations, using the average mass of the massive DPs in each model, rather than a single mass as adopted in \cite{Krivov18}. We observe some general agreement among our simulated models in following the behaviour described by equations (9) and (10) in \cite{Krivov18}, especially as the fraction of mass in the disc increases. However, we note that the theoretical models are only referential; thus, we made no attempt to fit our results to some particular parameter in such equations, for example, by varying the value of the numerical scaling factor $C_e$. 

In the following sections, we will consider the aspect ratio of the discs as represented by the RMS$_i$ of TPs, rather than that of DPs. This approach is taken because TPs are expected to accurately reflect the behaviour of dust, at least that dust not affected by non-gravitational forces. This method provides a more accurate proxy for real observations as opposed to large, massive perturbers.

\subsection{Grid of simulations: global evolution of RMS$_{e}$ and RMS$_{i}$}
\label{subsec:grid_sims_3_4}

\begin{figure*}
    \centering
    
    \includegraphics[width=0.33\textwidth]{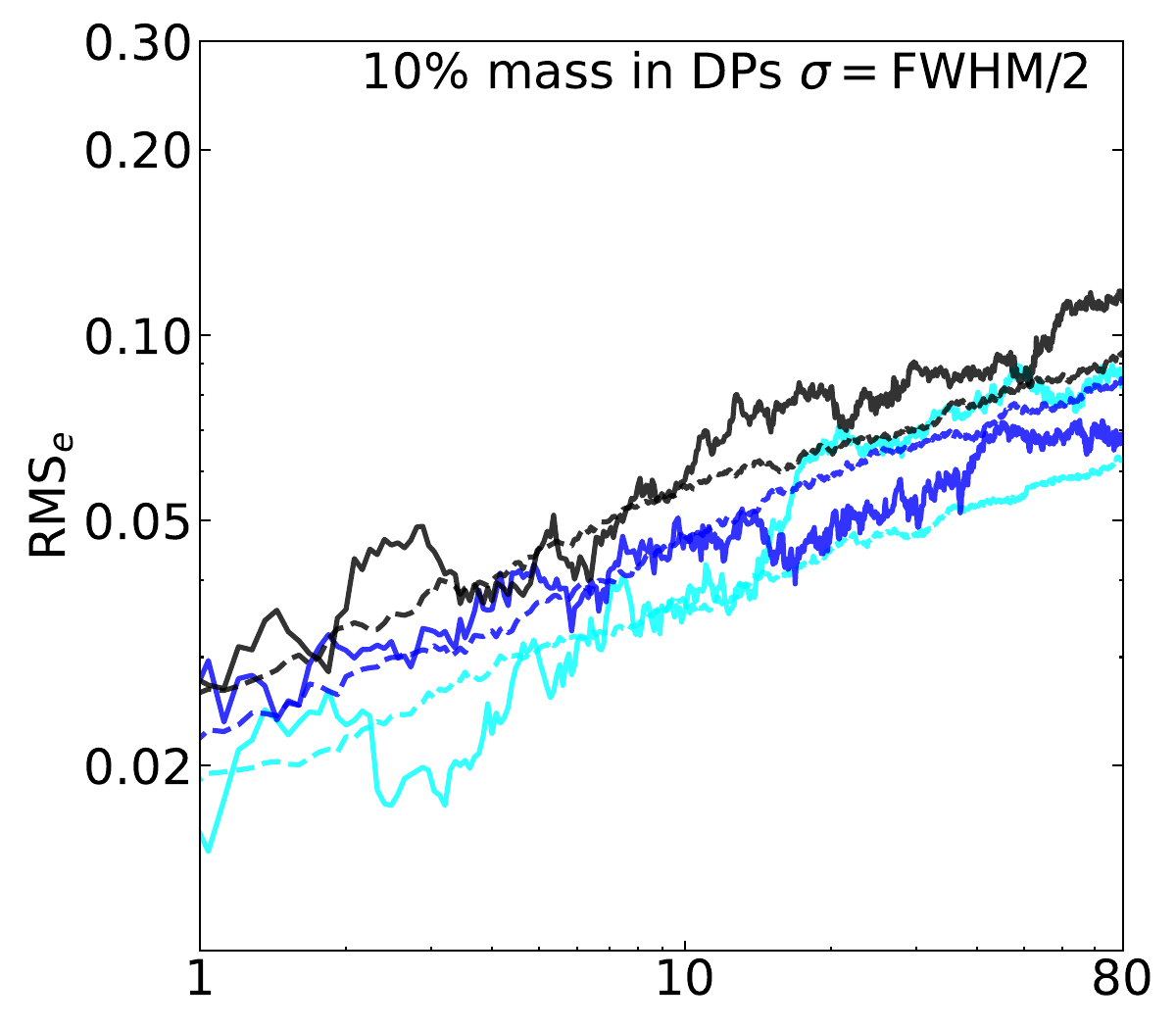}
    \includegraphics[width=0.33\textwidth]{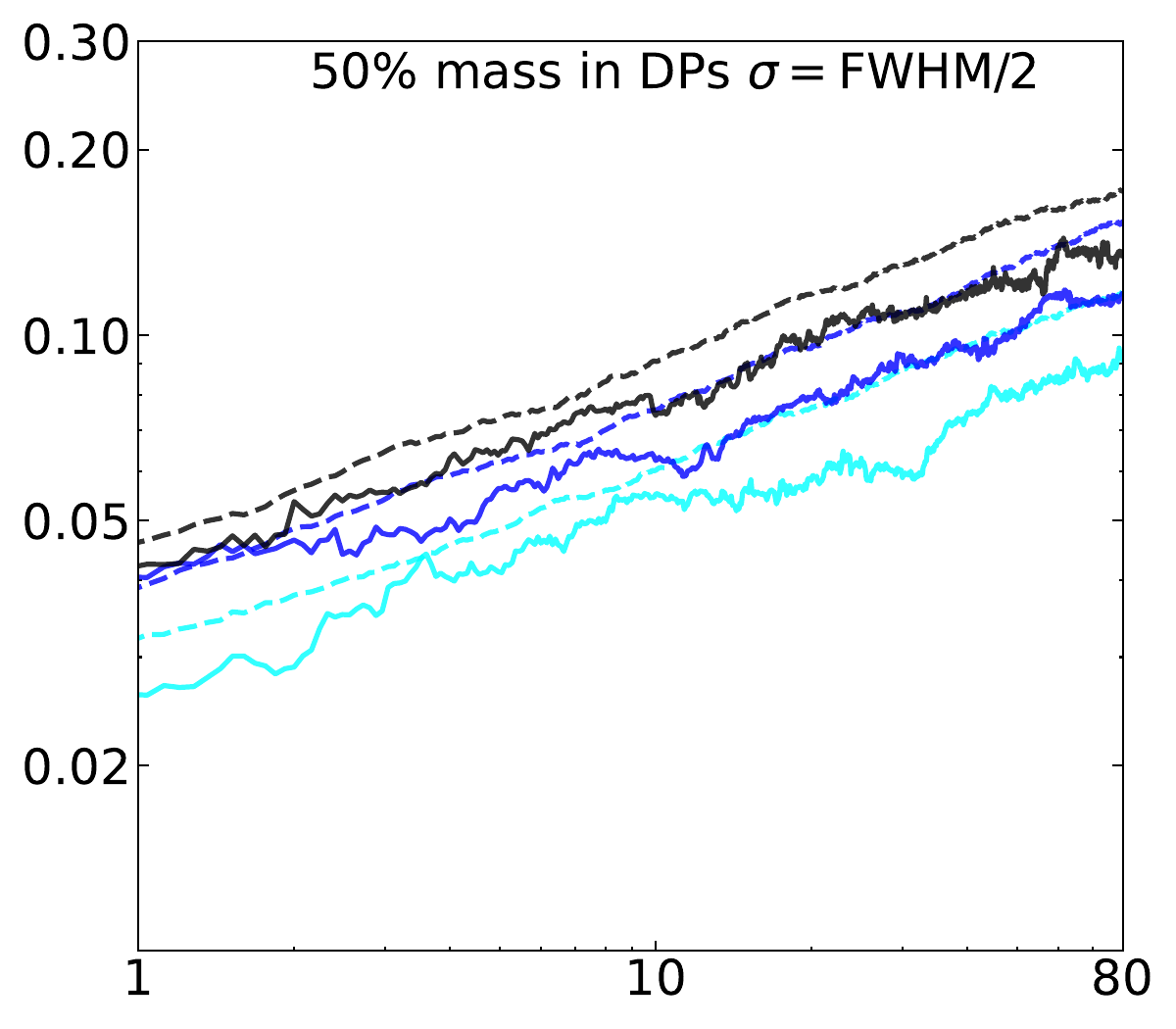}
    \includegraphics[width=0.33\textwidth]{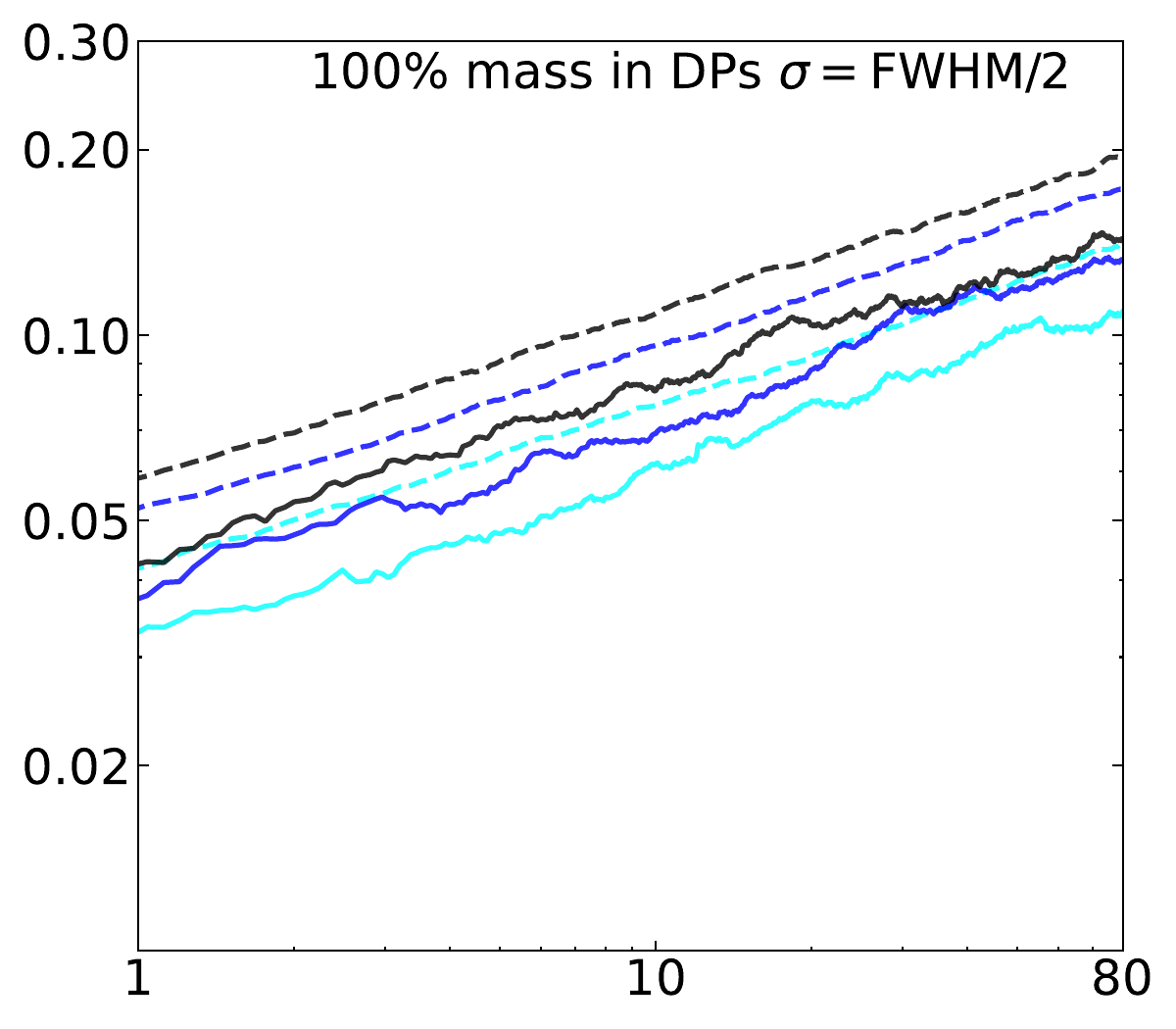}    \\

    \includegraphics[width=0.33\textwidth]{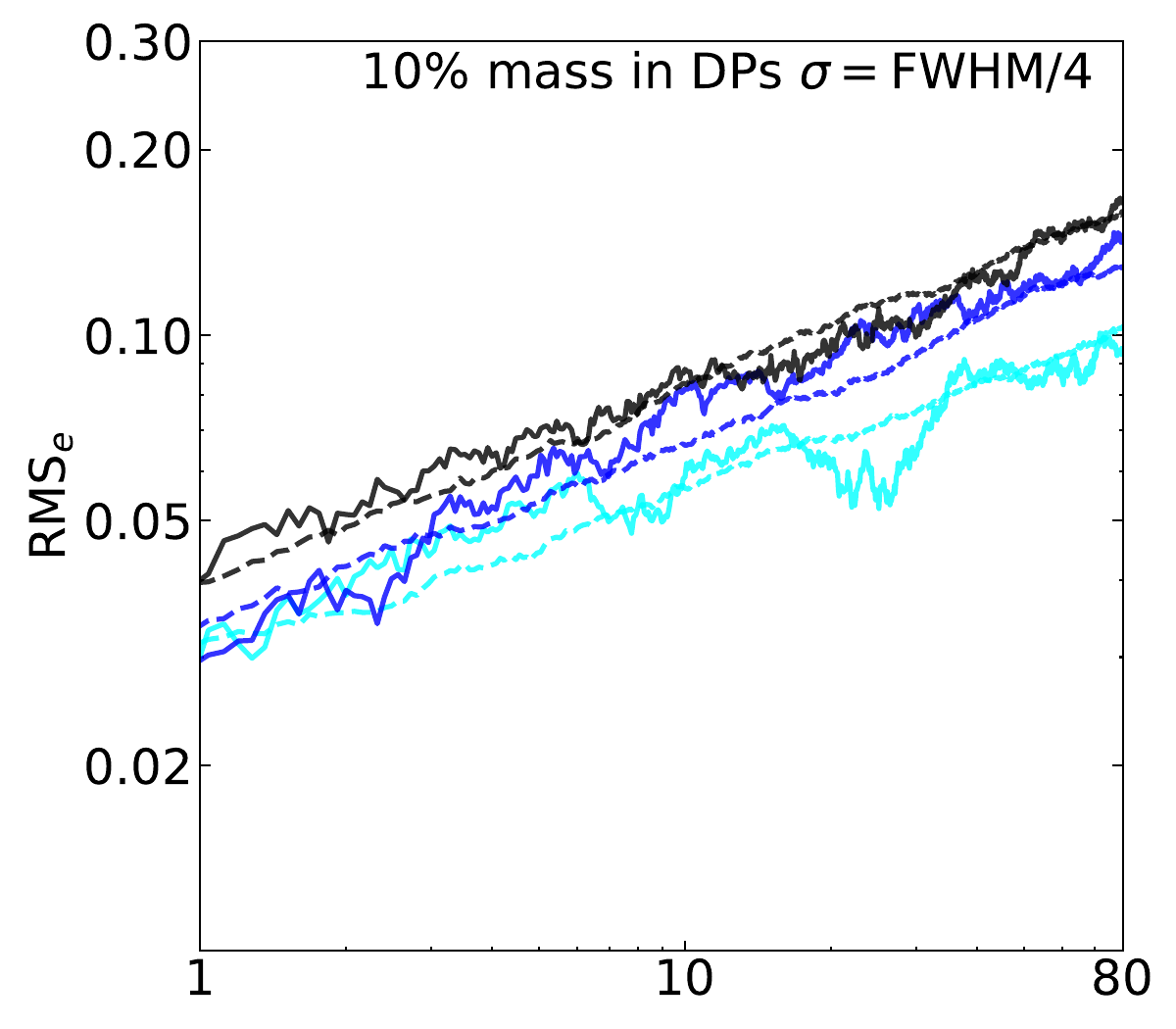}
    \includegraphics[width=0.33\textwidth]{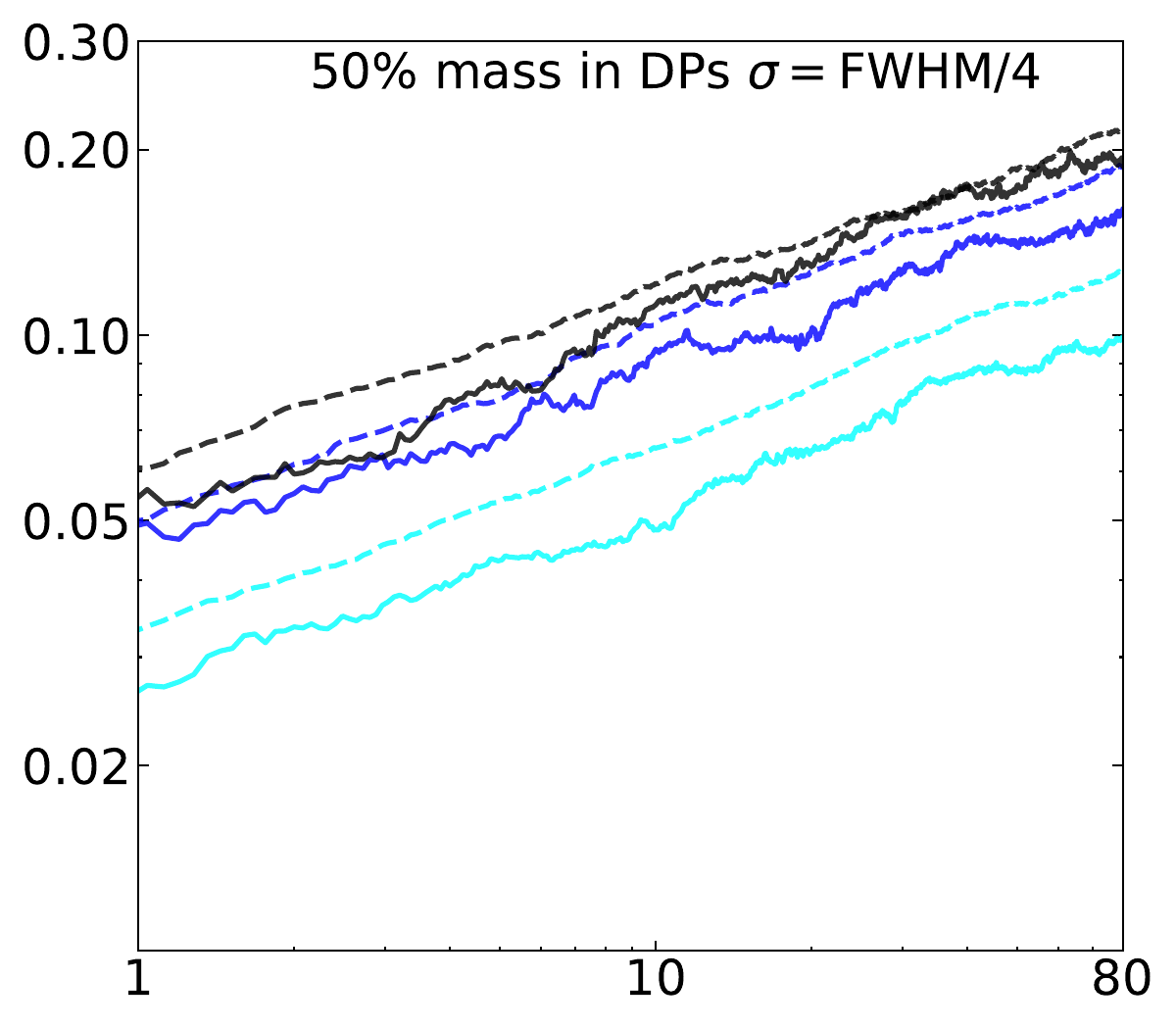}
    \includegraphics[width=0.33\textwidth]{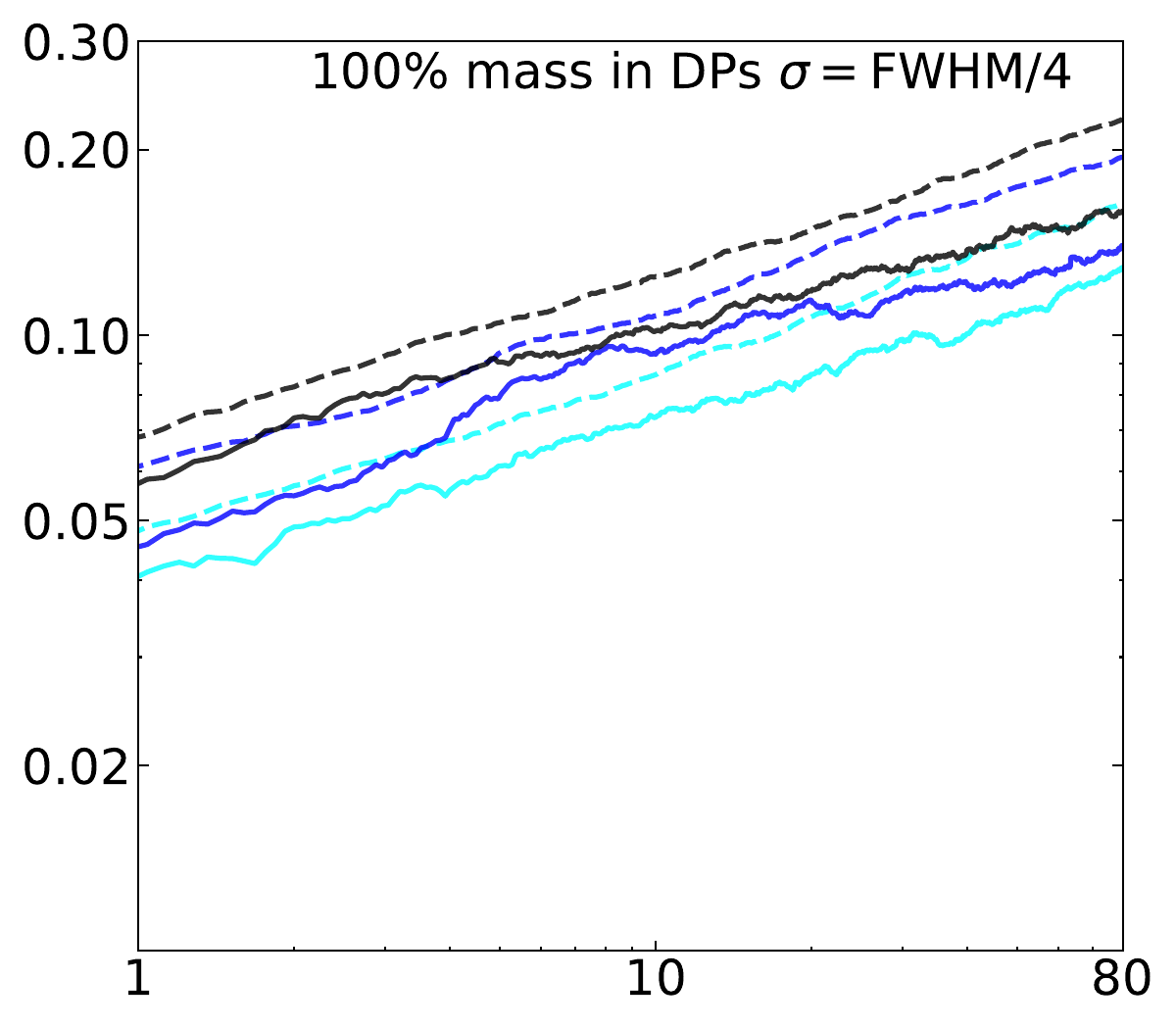}    \\

    \includegraphics[width=0.33\textwidth]{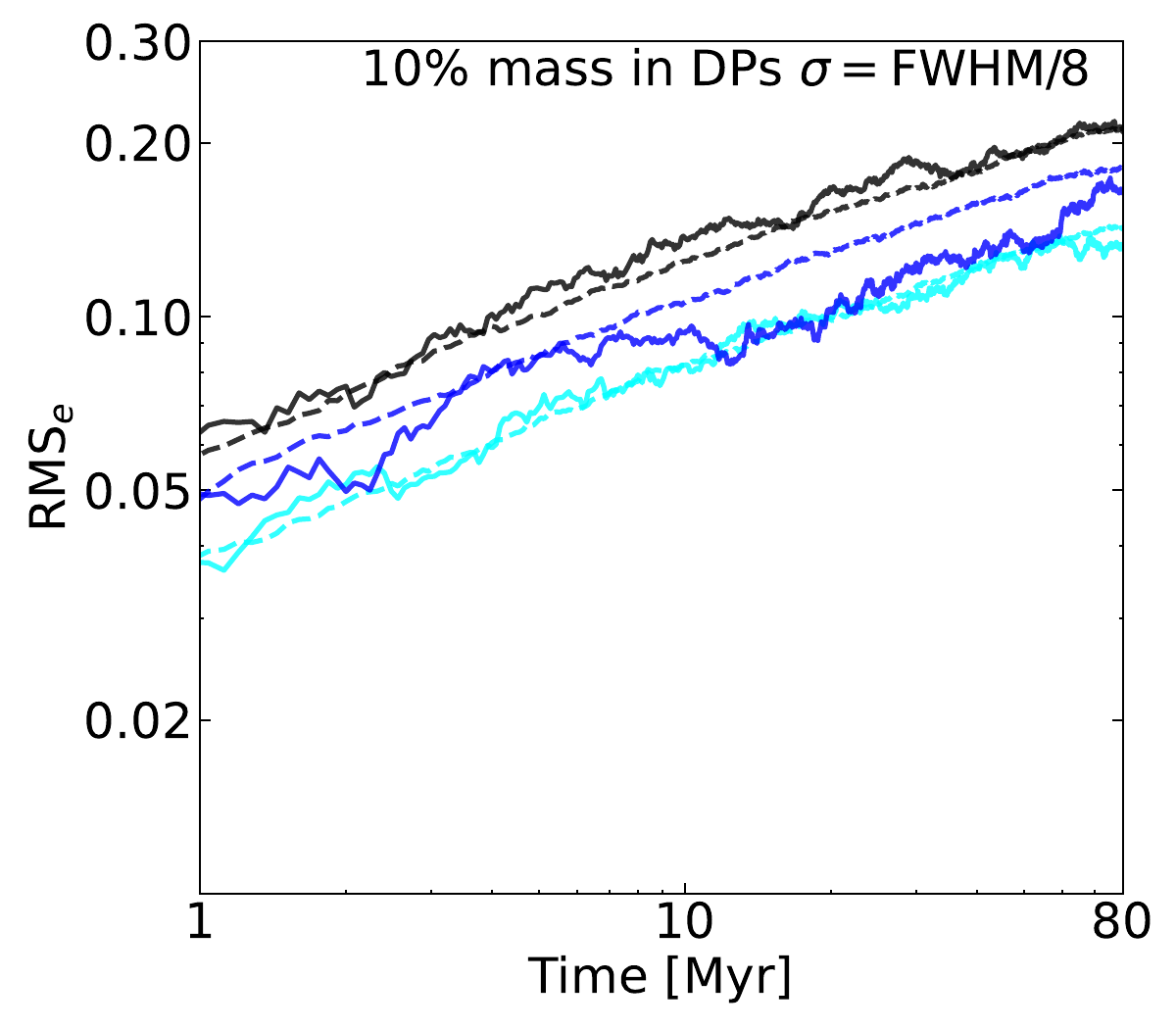}
    \includegraphics[width=0.33\textwidth]{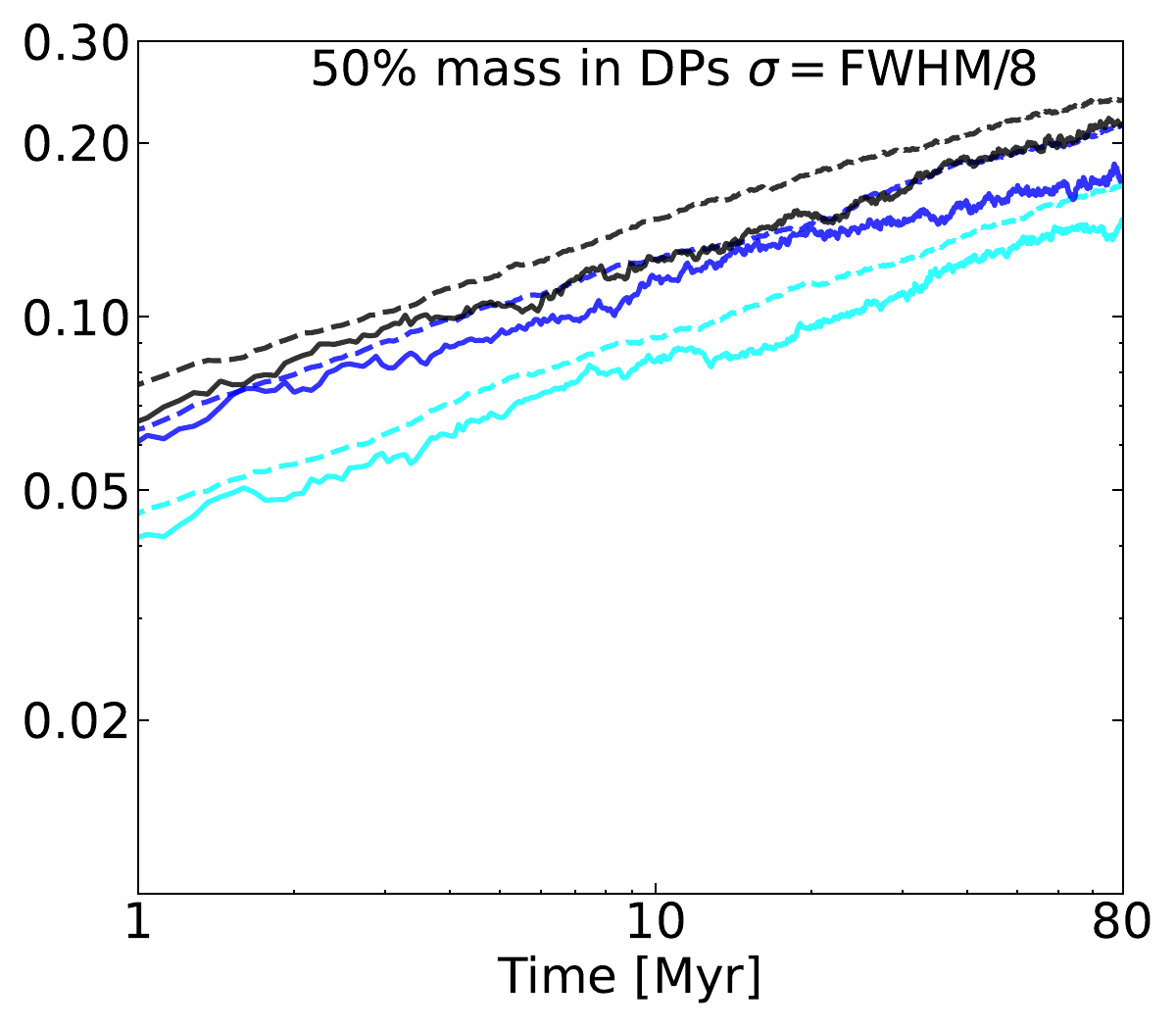}
    \includegraphics[width=0.33\textwidth]{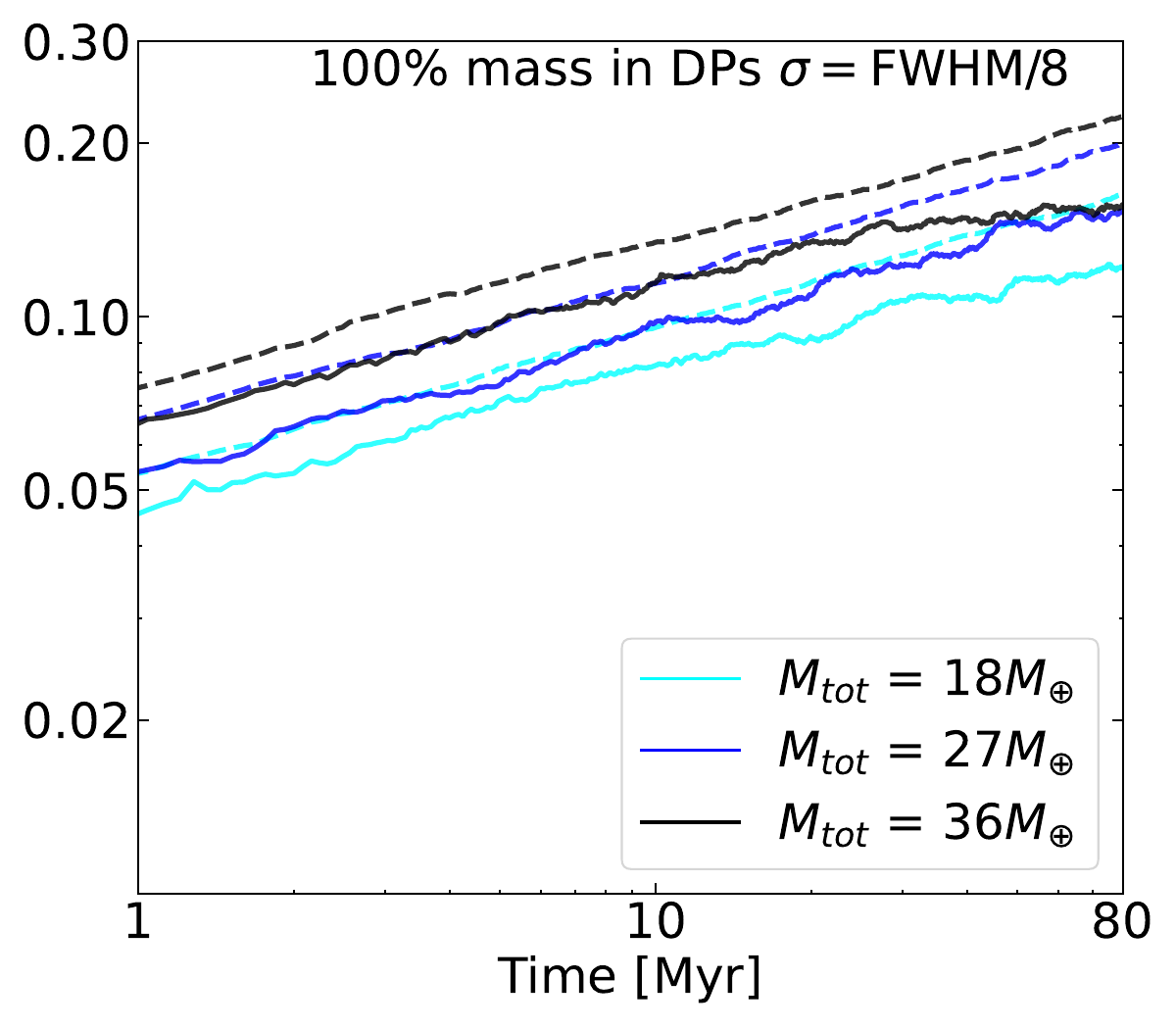}    \\
    
    \caption{Evolution of RMS$_{e}$ as a function of time for mass fractions in DPs of 10\% (left), 50\% (middle), and 100\% (right) and initial planetesimal belt widths of $\mathrm{FWHM}/2$ (top), $\mathrm{FWHM}/4$ (middle), and $\mathrm{FWHM}/8$ (bottom). Colours denote total disc mass of 18 (cyan), 27 (blue), and 36 (black) $M_{\oplus}$. Dashed lines denote RMS$_{e}$ of TPs, whilst solid lines denote RMS$_{e}$ for DPs.  See the text (Section \ref{subsec:grid_sims_3_4}) for further details.
    }
    \label{fig:rms_e}
\end{figure*}

\begin{figure*}
    \centering
    
    \includegraphics[width=0.33\textwidth]{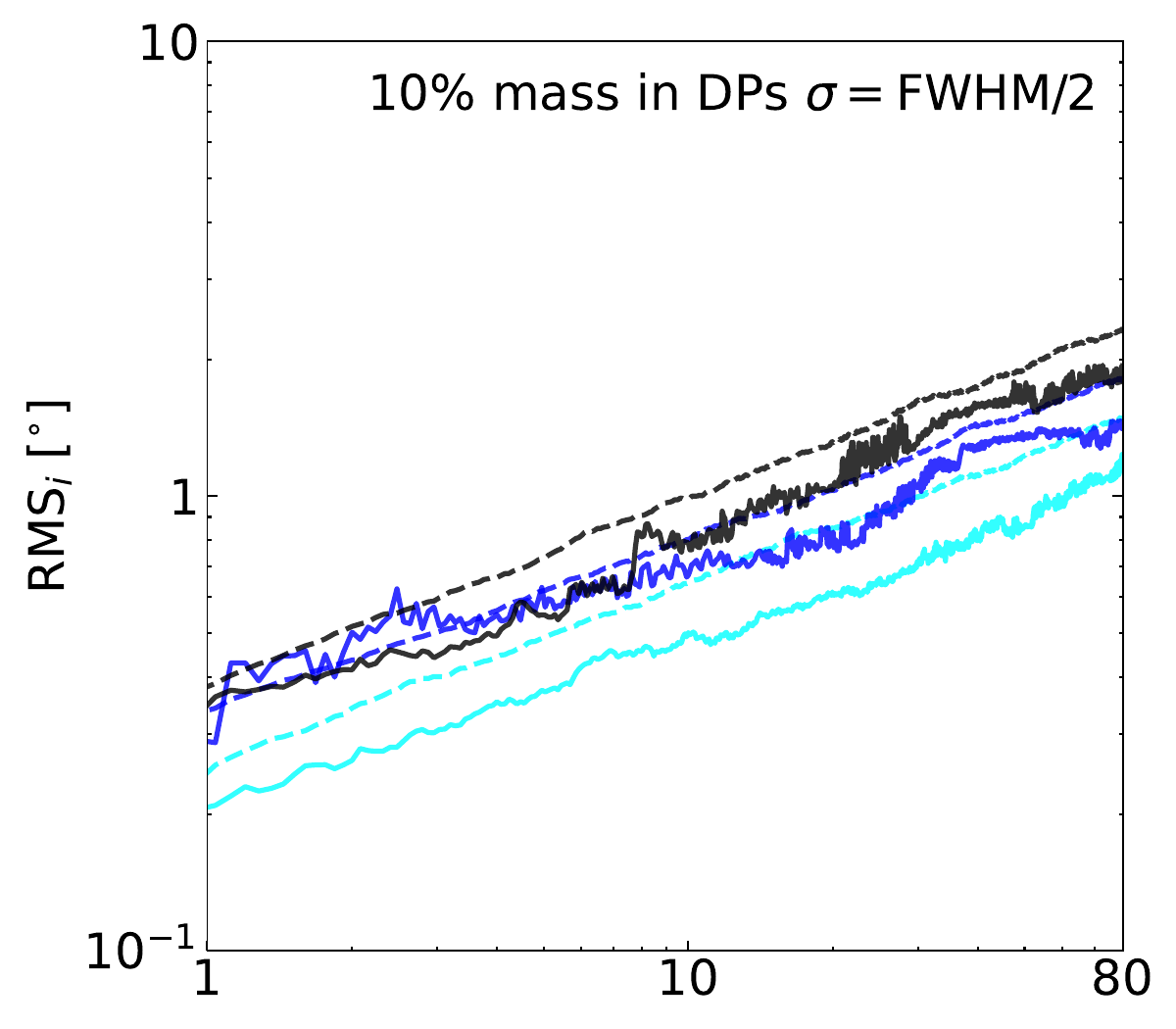}
    \includegraphics[width=0.33\textwidth]{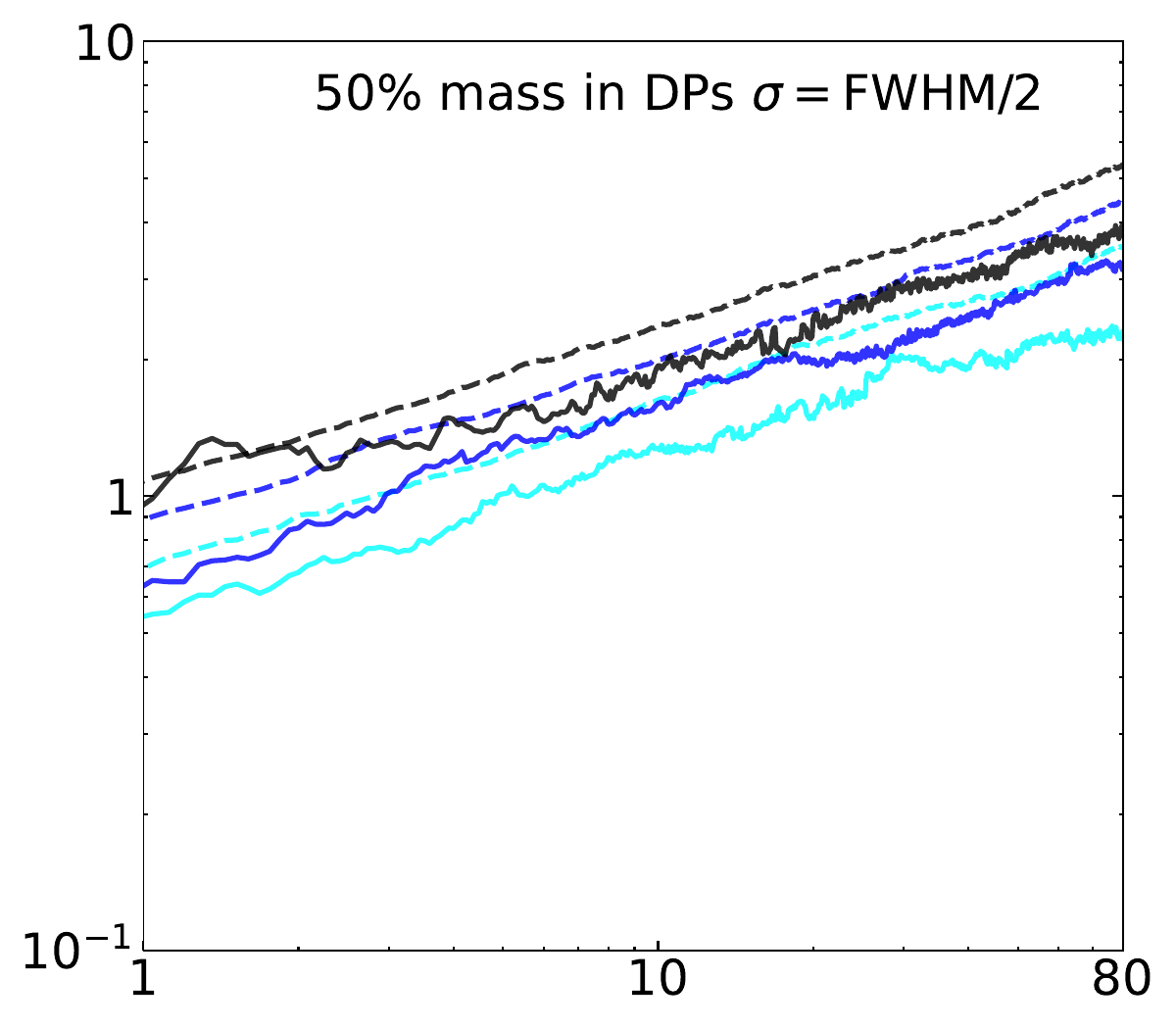}
    \includegraphics[width=0.33\textwidth]{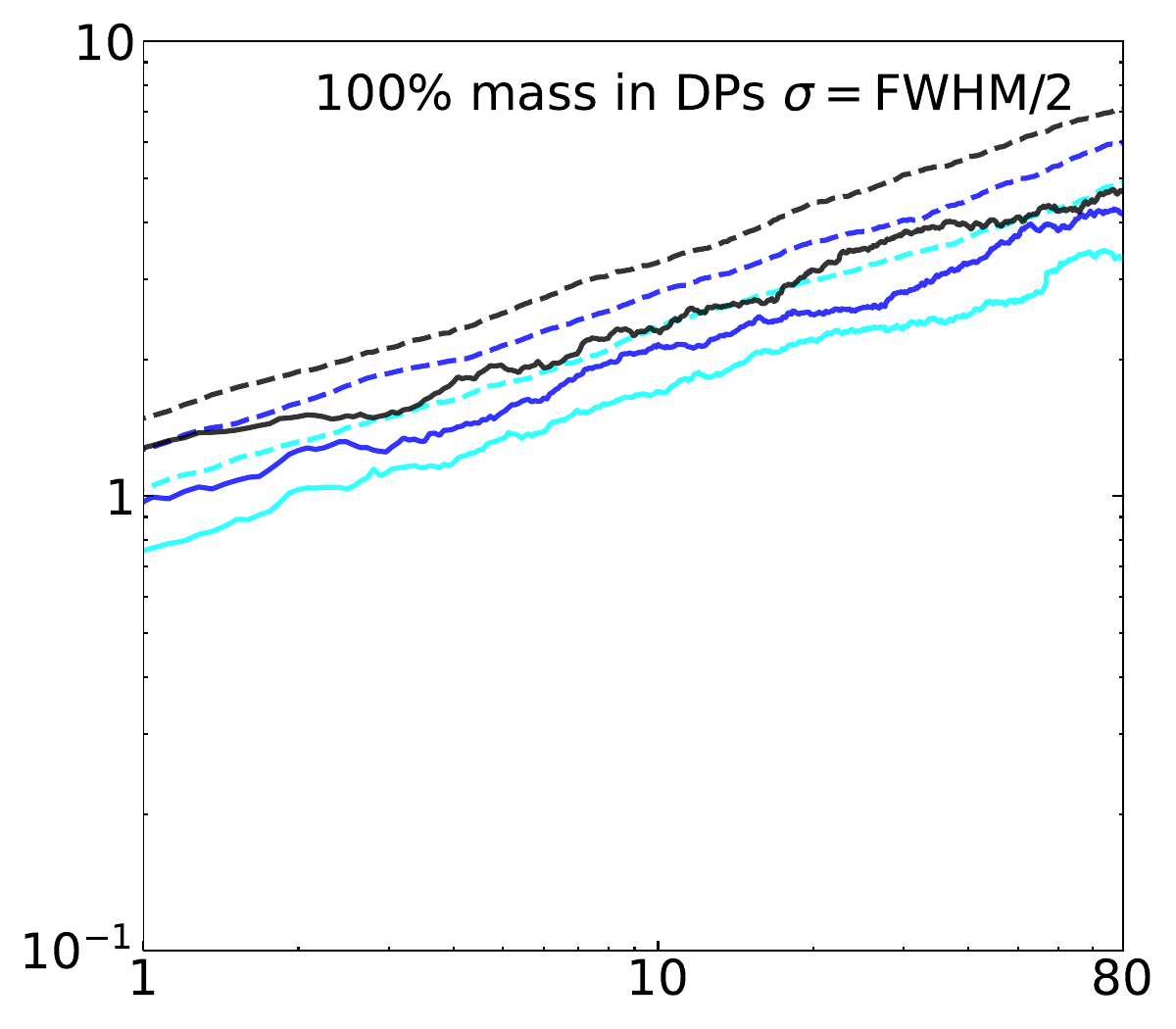}    \\

    \includegraphics[width=0.33\textwidth]{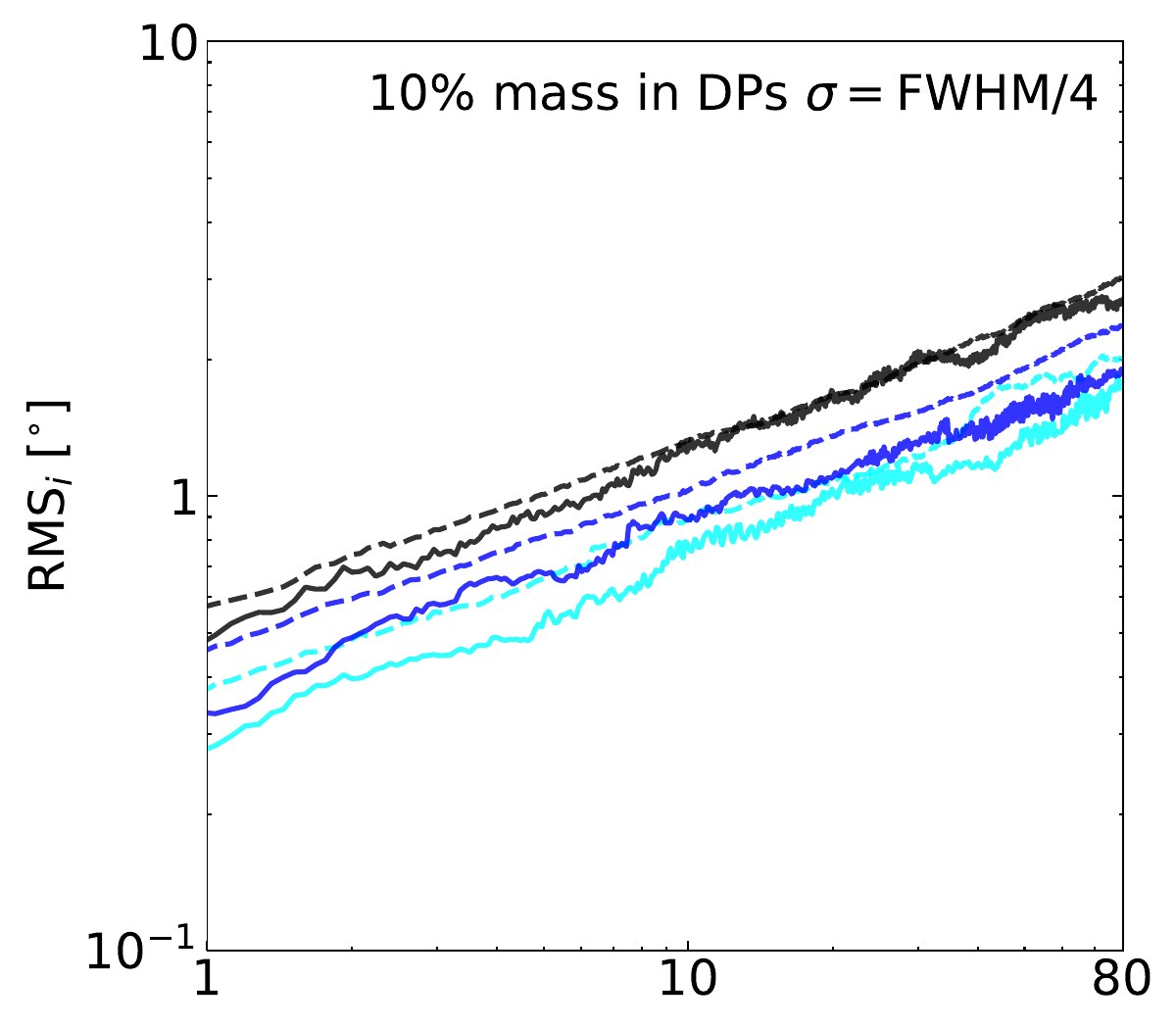}
    \includegraphics[width=0.33\textwidth]{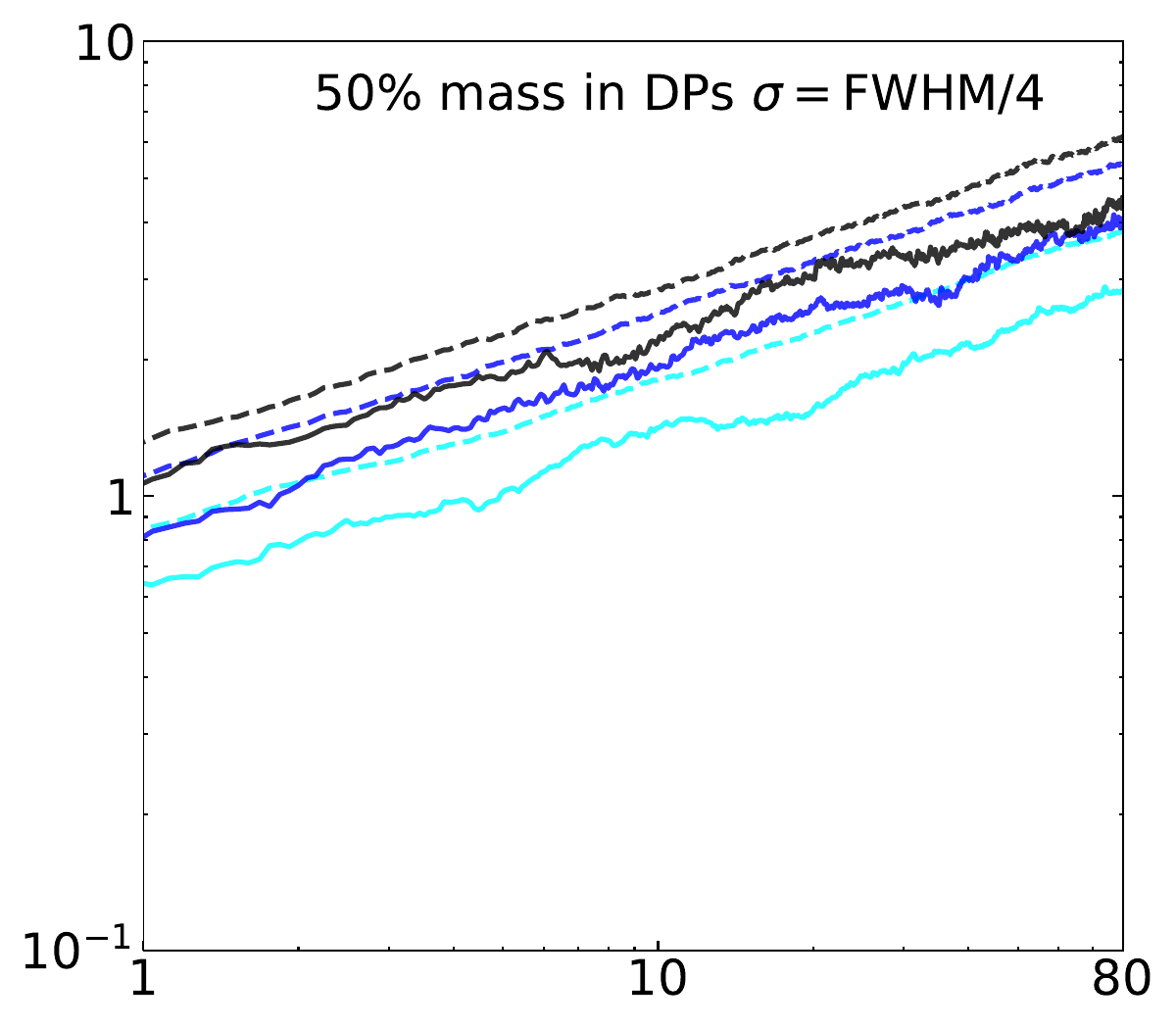}
    \includegraphics[width=0.33\textwidth]{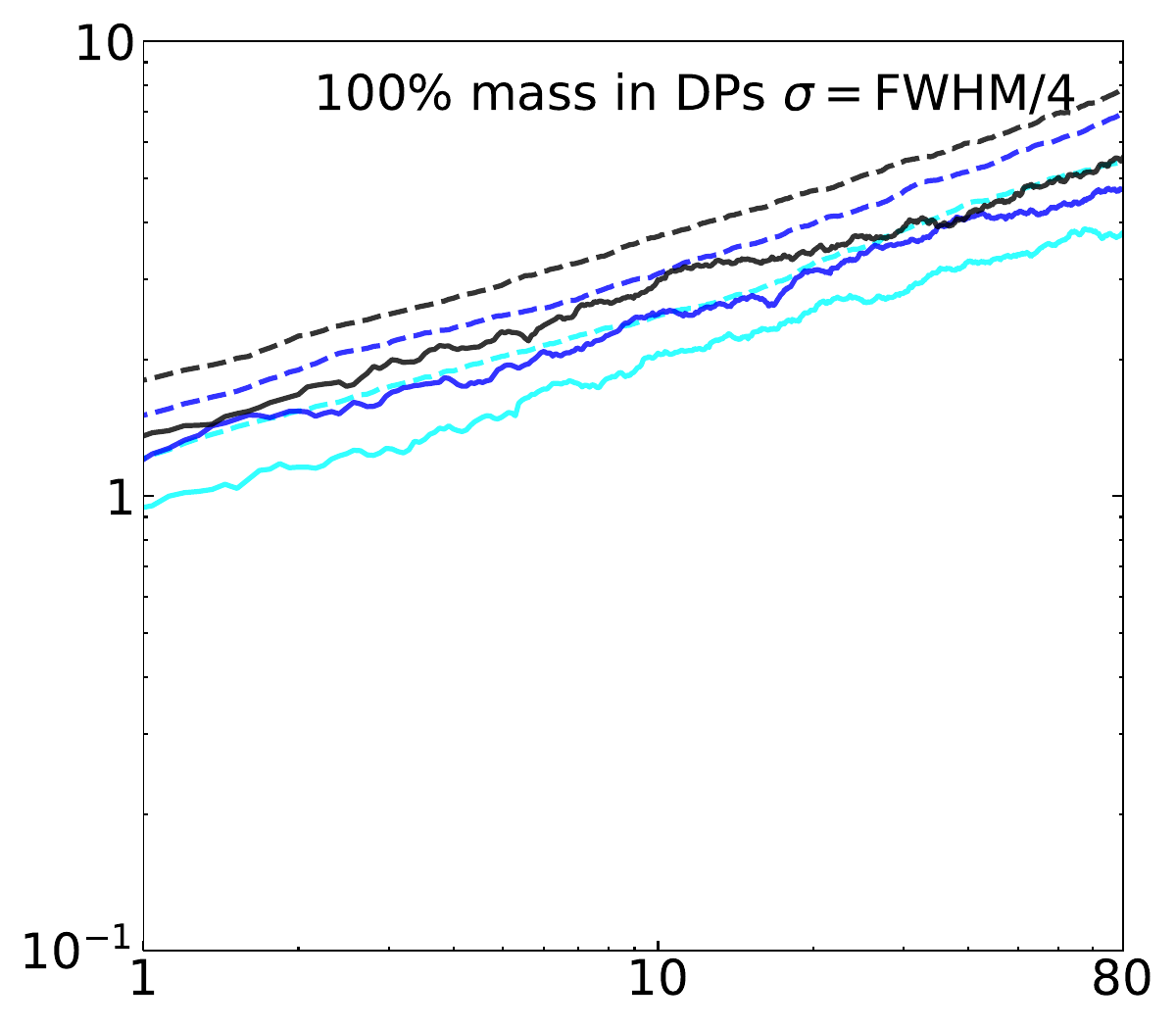}    \\

    \includegraphics[width=0.33\textwidth]{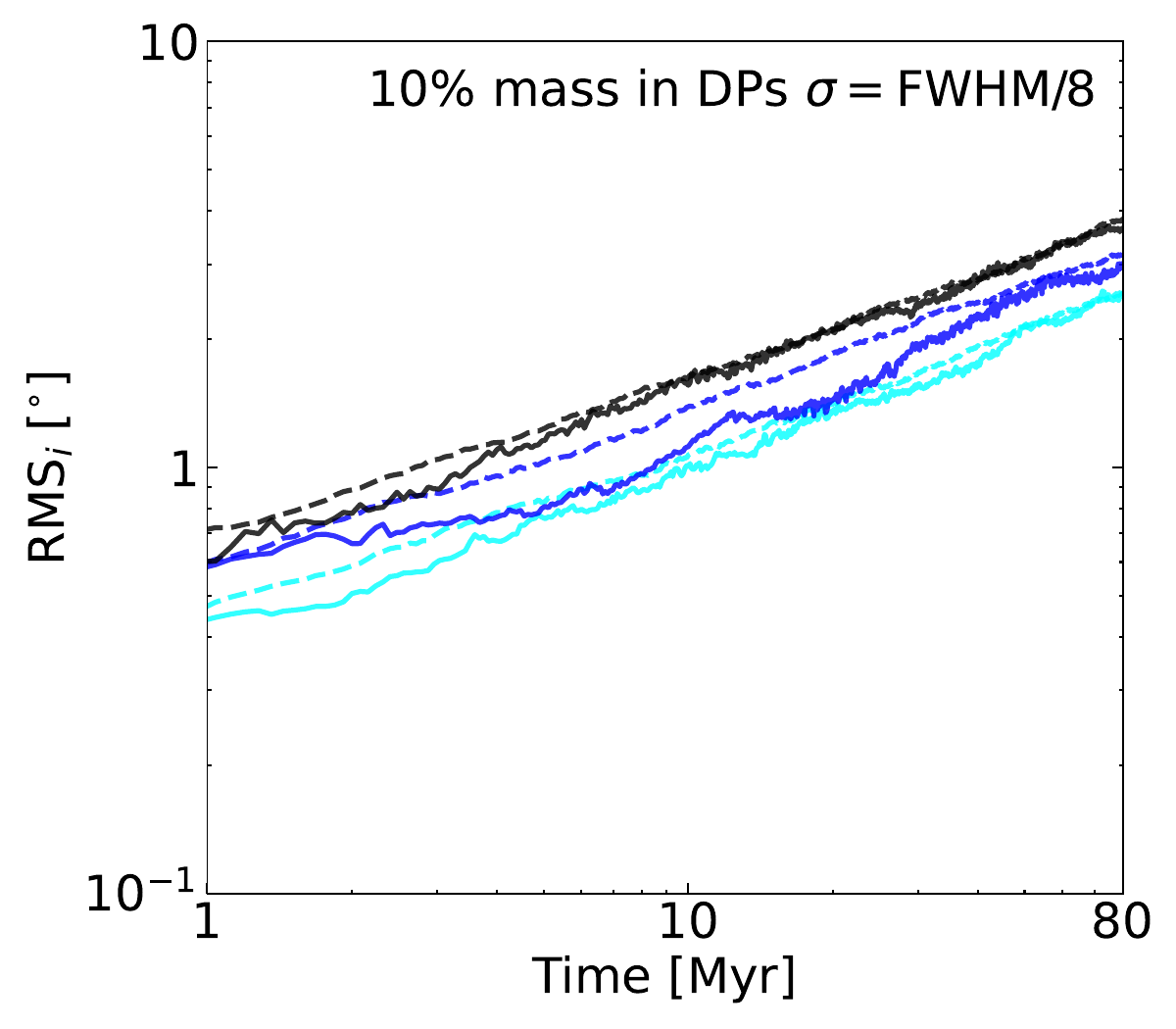}
    \includegraphics[width=0.33\textwidth]{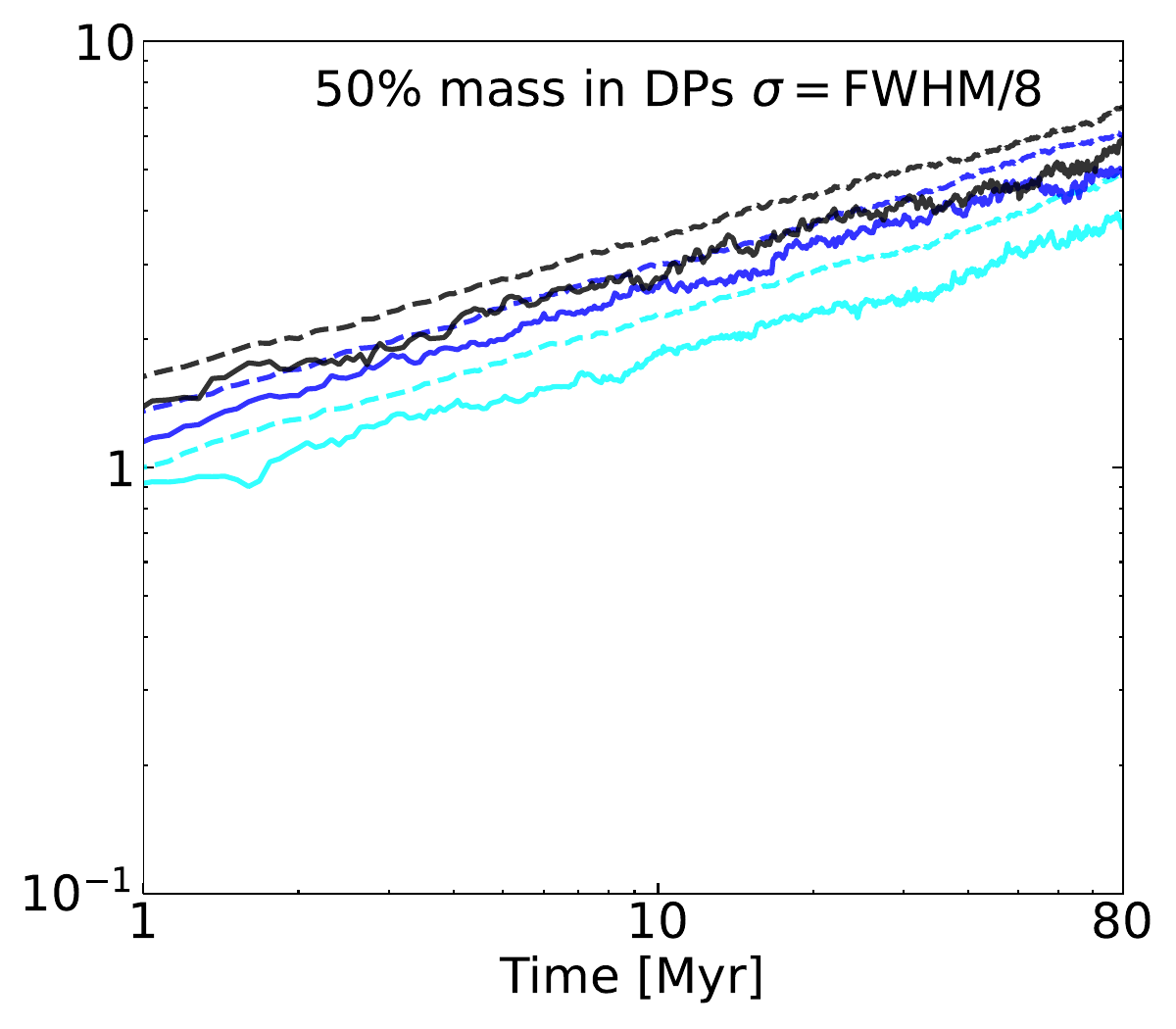}
    \includegraphics[width=0.33\textwidth]{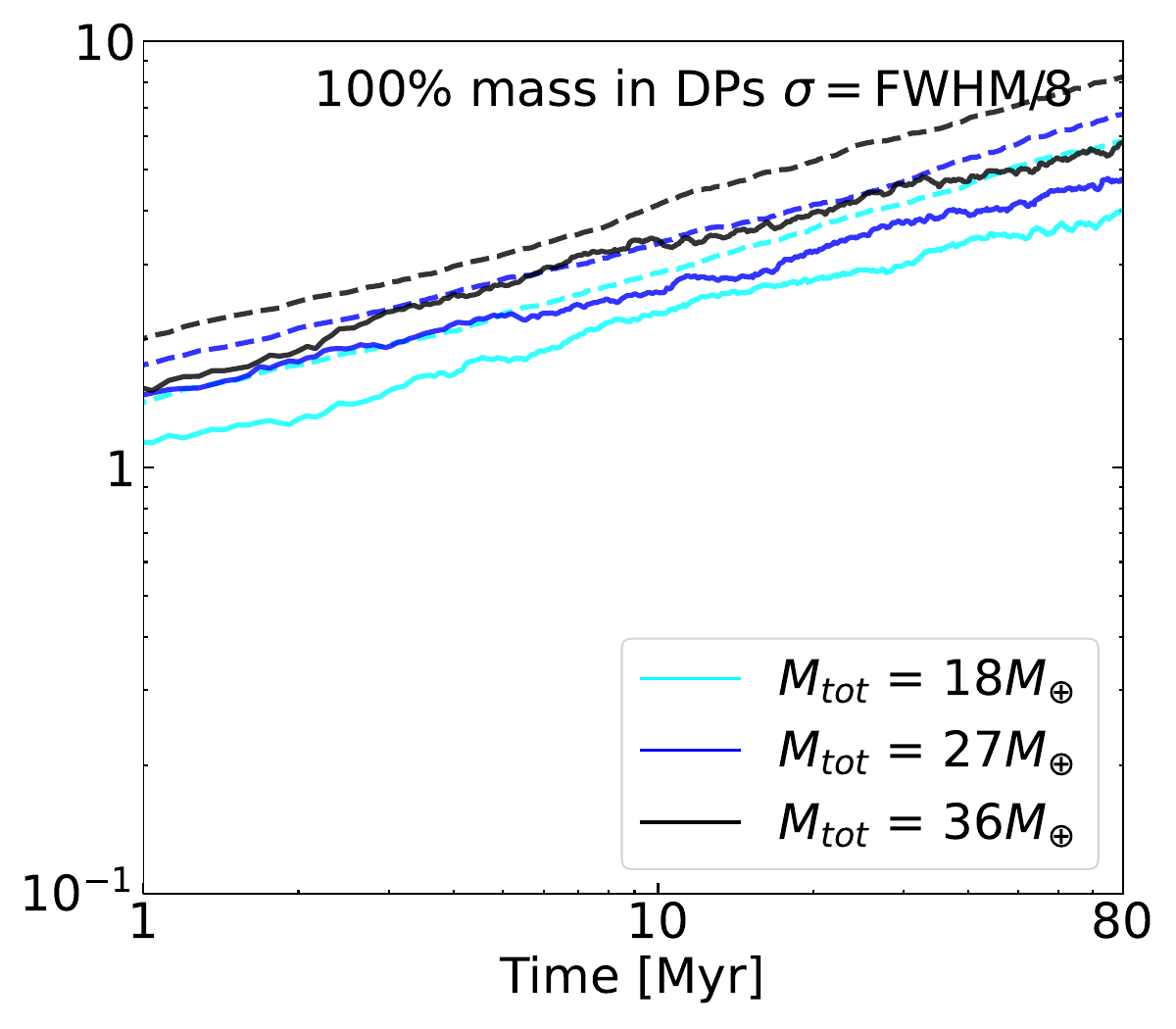}    \\
    
    \caption{Evolution of RMS$_{i}$, presented in the same manner as Figure \ref{fig:rms_e}. See the text (Section \ref{subsec:grid_sims_3_4}) for further details.}
    \label{fig:rms_i}
\end{figure*}

Having detailed the evolution of a system with mass 27 M$_\oplus$ and a disc width of $\mathrm{FWHM}/4$, we now present the results from our grid of simulations, which encompasses three different total masses, three mass fractions of planetesimals, and three different initial disc widths. The results for RMS$_e$ and RMS$_i$ are summarized in Figures \ref{fig:rms_e} and \ref{fig:rms_i}, respectively.

In Figure \ref{fig:rms_e}, the panels show the evolution of the RMS$_e$ for DPs in solid lines, and TPs in dashed lines. The columns correspond to an increasing fraction of the disc mass in DPs (10, 50, 100\%; left to right) and the rows correspond to increasingly narrower belt widths (FWHM/2, /4, /8; top to bottom). Generally, while GPs are equally efficient in stirring DPs and TPs, DPs are more efficient in stirring TPs than other DPs. In the left column of Fig. \ref{fig:rms_e}, DPs are not massive enough to be the dominant source of stirring, but instead boost the effects of the GP. The difference between the eccentricity of the DPs and TPs in these simulations is only stochastic. On the other hand, in the center and right columns, the DPs are the dominant source of stirring; thus, the stirring of the TPs is always more efficient than that of the DPs. For the simulations with 100\% mass in DPs, the eccentricity stirring of the DPs slows down after approximately 30 Myr, with this effect being more noticeable for the $\sigma=\mathrm{FWHM}/8$ panel. 

Figure \ref{fig:rms_i} shows the evolution of the RMS$_i$ for DPs and TPs in the same way as Figure \ref{fig:rms_e} does for eccentricity. As with eccentricities, the difference in excitation levels between DPs and TPs increases with higher disc mass fractions in DPs, with TPs becoming more excited than DPs.
As was observed for eccentricity, in Figure \ref{fig:rms_i} we also see some cases where the DPs inclination stirring slows down; however, it appears to start a bit earlier, around 25 Myr, and occurs over a broader range of parameter space, i.e., when the mass is at least 50\% in DPs and $\sigma \leq \mathrm{FWHM}/4$. These results suggest that with the system configurations used in our simulations, we are approaching the saturation limit for both elements at 80 Myr of integration. 

\begin{figure*}
    \includegraphics[width=\textwidth]{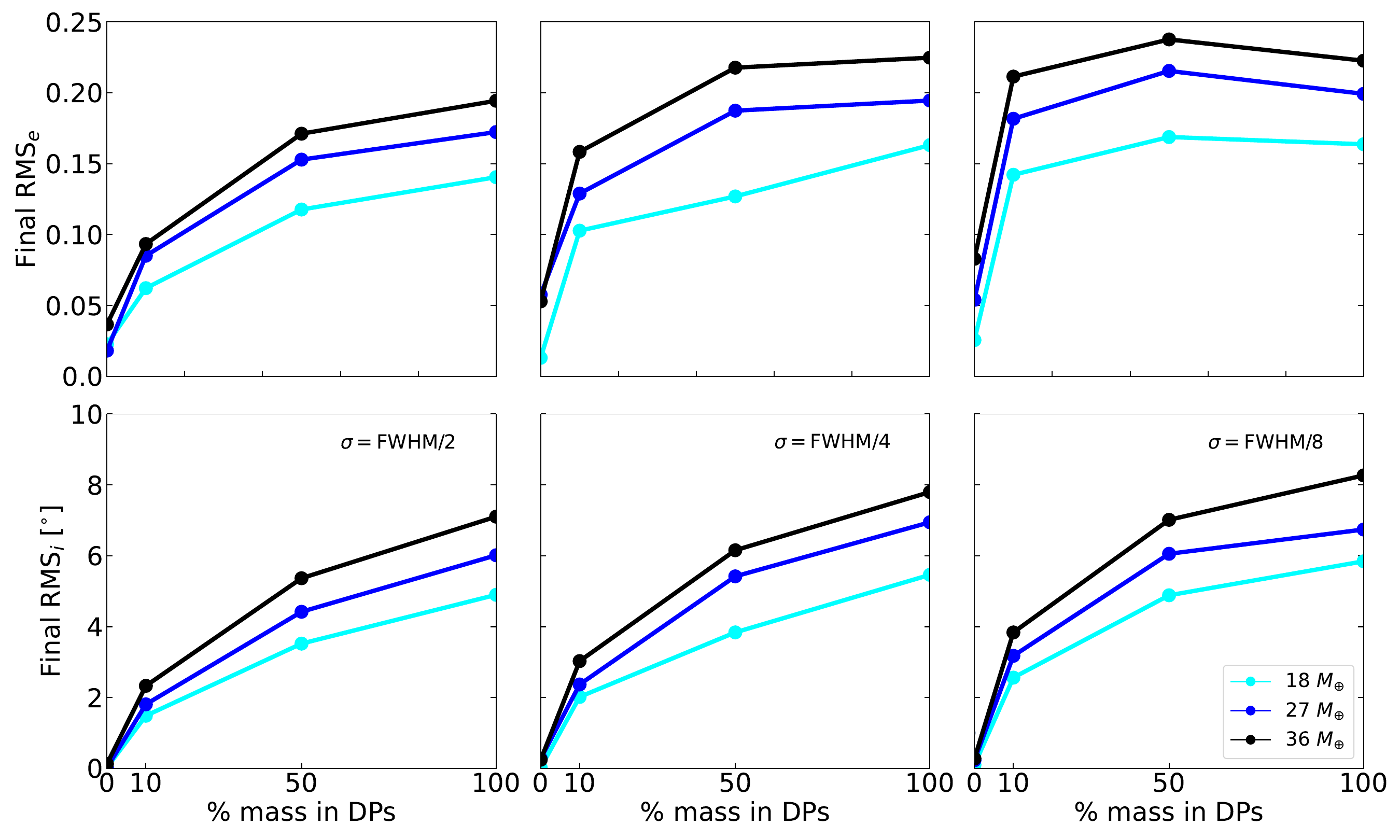}
    \caption{Final ($t = 80~$Myr) RMS$_{e}$ (top) and RMS$_{i}$ (bottom) values as a function of the disc mass in DPs for models with total masses of 18 (cyan), 27 (blue), and 36 (black) $M_{\oplus}$ and fractions of 10\%, 50\%, and 100\% in dwarf planets (DPs). The left, middle, and right panels show results for initial belt widths $\sigma$, factors of 2, 4, and 8 narrower than the width inferred from observations. See the text (Section \ref{subsec:grid_sims_3_4}) for further details.}
    \label{fig:rms_final}
\end{figure*}

Although Figures \ref{fig:rms_e} and \ref{fig:rms_i} provide all the relevant information about the stirring evolution of our models during the 80 Myr of integration, comparing the final state of the discs across different models is difficult. To compare this final state more effectively, in Figure \ref{fig:rms_final} we plot the final value of both the RMS$_e$ and RMS$_i$ as a function of the percentage of the mass in the disc, for all considered models.
The trend in Figure \ref{fig:rms_final} appears quite straightforward: for discs with a given initial radial width, larger total system mass leads to 
higher disc excitation after 80 Myrs. However, there are some important subtleties worth discussing. 
The singular exception to the above statement is when all the mass is in the giant planet (0\% case), in which case some lower-mass systems reach higher values; this is most likely due to statistical noise. The noise is statistical in the sense that it relates to the initial distribution of test particles in the model, since the excitation in eccentricities/inclinations is driven solely by their proximity to MMRs in this case.

We also notice that, in nearly all cases, the final values of both RMS$_e$ and RMS$_i$ are higher when all the mass is in the disc (i.e., in the equivalent self-stirring scenario). While we might conclude that the self-stirring model is more efficient than the mixed-stirring model, this is challenged by the presence of a turnover in the RMS$_{e}$ curves for cases where the disc is initially narrow in terms of radial width. Specifically, for $\sigma=\mathrm{FWHM}/8$, the heating of the eccentricity for all three values of $M_{\rm tot}$ peaks when the DP mass is at 50\%.

The curves in Figure \ref{fig:rms_final} demonstrate that models incorporating a combination of effects, i.e., a giant planet and a massive disc, are sufficiently effective in exciting disc particles, eliminating the need for a purely self-stirring scenario. 
This is particularly relevant because self-stirring models encounter difficulties explaining systems without resorting to disc masses that are so large ($\geq1\,000~M_{\oplus}$) that the amount of solids in their progenitor protoplanetary discs would be unrealistically large when assuming typical dust-to-gas ratios \citep{Manara18,WyattKrivov21}.

To put it differently, the previous results indicate that, while self-stirring is critical and can even explain all the disc's vertical heating, it is also feasible to assign some of the mass to an internal GP. This configuration has the additional advantage of creating a sharp inner boundary to the disc \citep[e.g. ][]{Pearce24}, similar to what is observed in HD~16743. Finally, in nature, the heating of some objects will occur through planetary perturbations, others through DP self-stirring, and others will have mixed stirring. Our work presents some models that explore the differences between all three stirring models. In the case of HD~16743, we conclude, based on Figures \ref{fig:rms_e}--\ref{fig:rms_final}, that a mixed model can achieve stirring levels comparable to those of a purely self-stirring model, but with only half the mass in the disc, thanks to the presence of a relatively small giant planet.

Finally, we briefly summarise the evolution of the GP throughout the simulations presented here. We initially place the GP separated from the inner edge of the disc by 3 Hill radii. This is close enough to induce stirring (within 5 Hill radii) but far enough to avoid the belt being disrupted at the first timestep. Ignoring the simulations with 100\% of disc mass in DPs (as the GP effectively becomes a TP), in all cases the GP undergoes only minor excitation over the duration of the simulations with a $< 1\%$ change in $a_{\rm GP}$, $e_{\rm fin, GP} \leq 0.01$, and $i_{\rm fin, GP} \le 0.1\degr$. As such, a massive GP does not undergo significant dynamical evolution during the course of these simulations.

\section{Observational Modelling}
\label{sec:results_obs}

We begin by constructing surface brightness maps from the simulated TPs, following the method of \citet{Sefilian21} and \citet{Farhat23}, before converting these into synthetic observations. To this end, each simulated particle is assigned a constant mass $m_i$ (i.e., $m_i = m_{i+1}$ for all $i = 1, ..., 10^3$) and is smeared over its orbit by generating $N_c = 10^4$ clones. Each clone has a mass of $m_i/N_c$ and shares the parent planetesimal's orbital elements, but is given a randomly selected mean anomaly in the range $[0, 2\pi]$. This procedure not only improves the resolution of the resulting maps but also effectively transforms each planetesimal into a ring with a non-uniform linear density -- accounting for the fact that particles on eccentric orbits spend more time near apocentre than pericentre. To compute the disk's surface brightness, the resulting cloud of clones is first weighed by a factor of  $R^{-1/2}$, where $R = \sqrt{X^2+Y^2+Z^2}$, consistent with thermal emission in the Rayleigh-Jeans limit from large (i.e., $\gtrsim$mm-sized) particles under the assumption of optically thin, blackbody emission (in which case the temperature scales as $T \propto R^{-1/2} L_{\star}^{1/4}$). The weighted particle cloud is then projected onto the sky plane, accounting for the disc’s orientation, and binned into a grid with a pixel area of $0.25\times0.25$ au$^2$ to produce a projected surface brightness map.

In creating the maps of surface density/brightness, we ignore the contribution of the DPs. This is because the emission efficiency of particles drops at wavelengths longer than their physical sizes. At the same time, particles larger than $\sim $ cm remain invisible at any wavelength; simply because the cross-section that they carry is small for reasonable size distributions in debris discs \citep[i.e., $n(D) \propto D^{-q}$ with $q = 3.5$ for steady-state collision,][]{1984DraineLee}. 

\begin{figure*}
    \raggedright
    \includegraphics[width=0.33\linewidth,trim=0 -1.4cm 0 0]{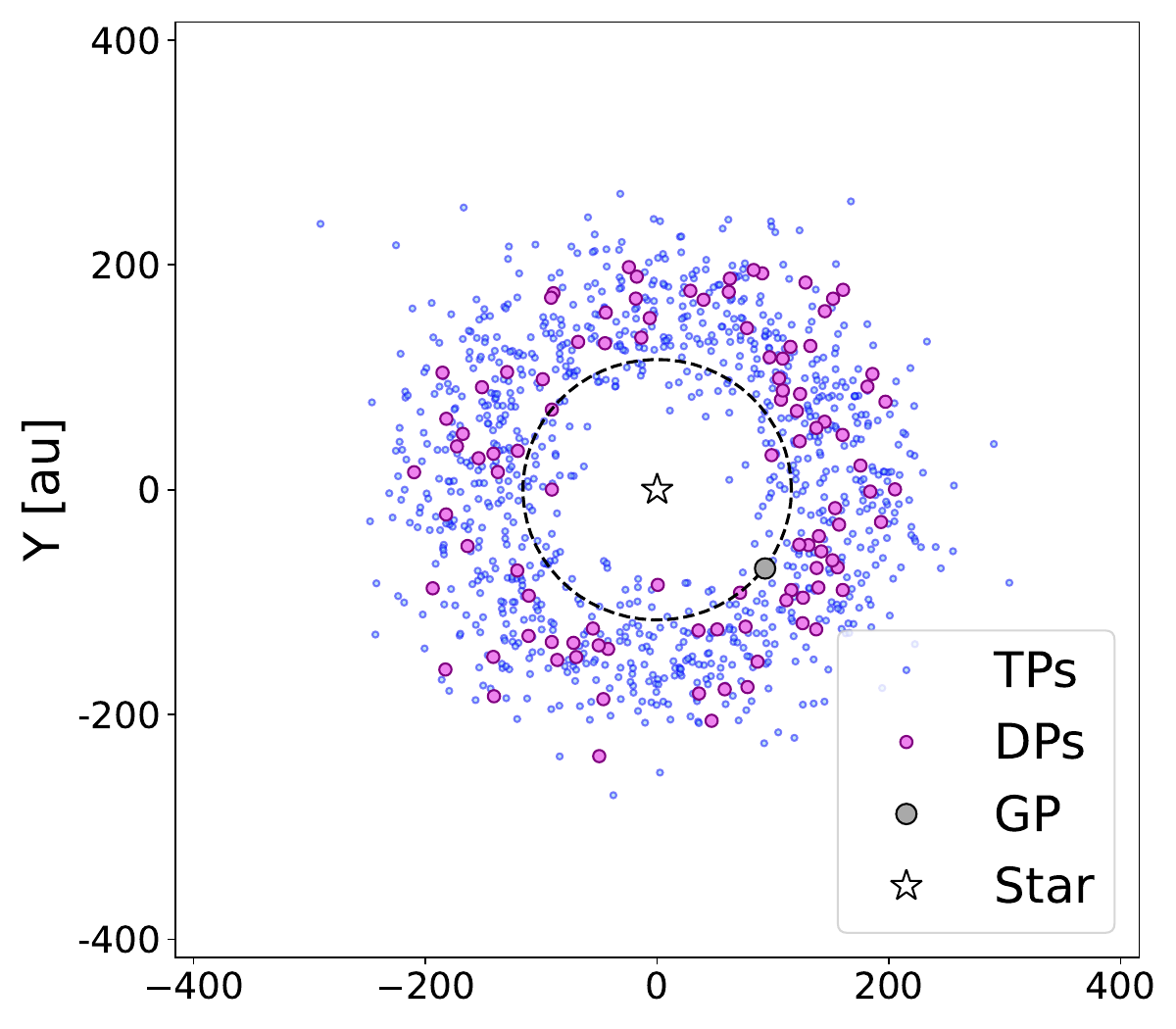}
    \includegraphics[width=0.33\linewidth,trim=0 2cm 0 0]{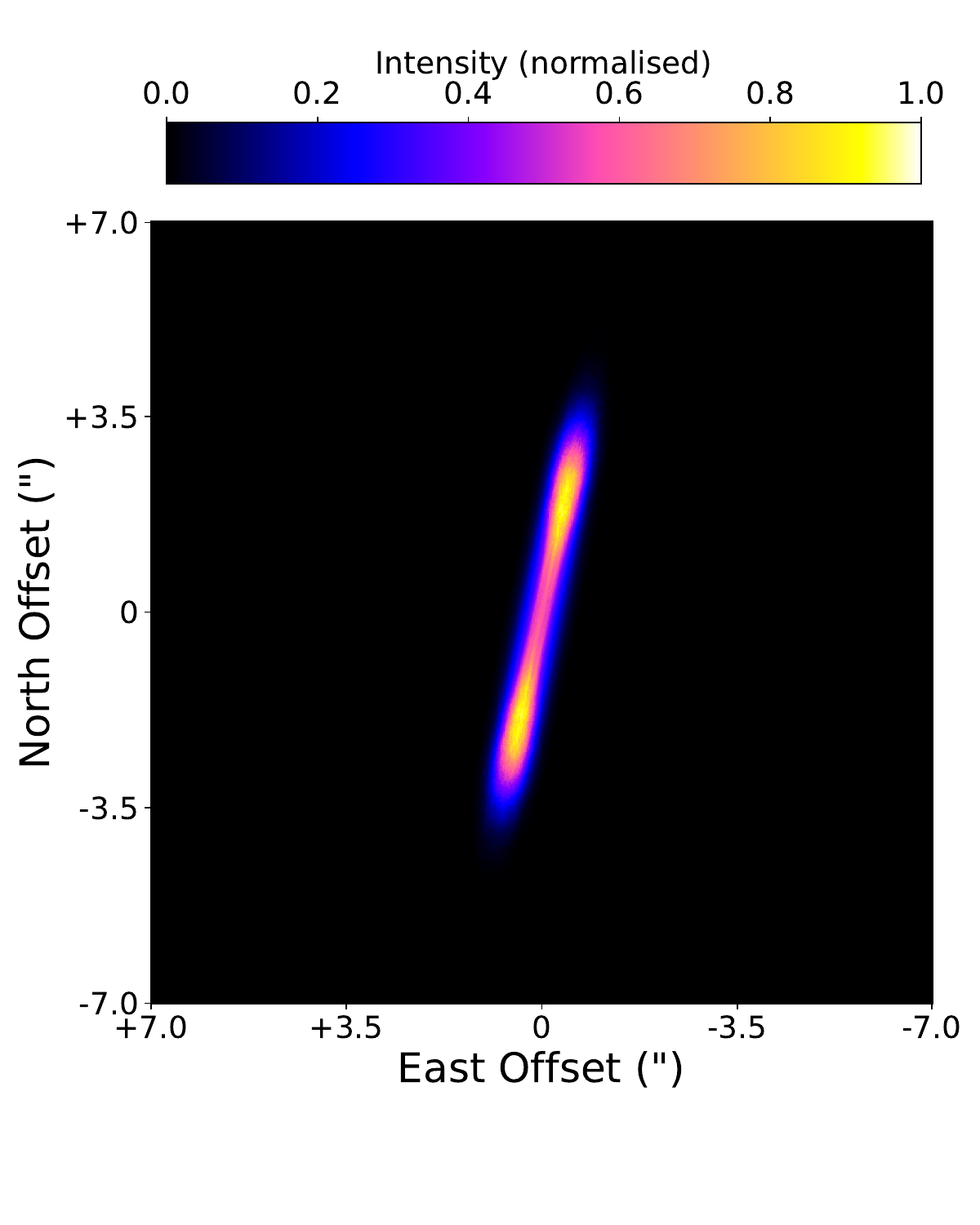}
    \includegraphics[width=0.33\linewidth,trim=0 2cm 0 0]{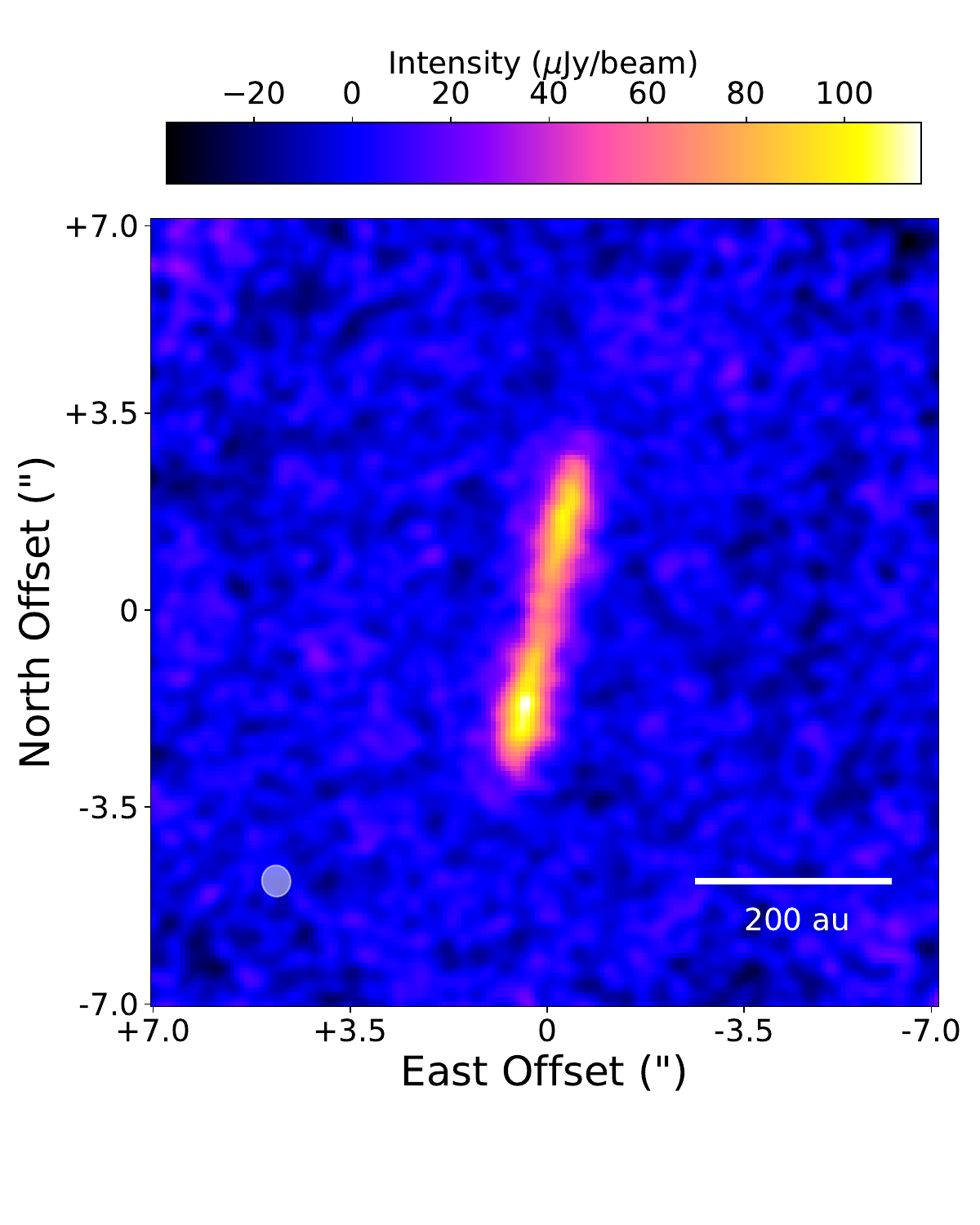}\\
    \includegraphics[width=0.33\linewidth,trim=0 0cm 0 0]{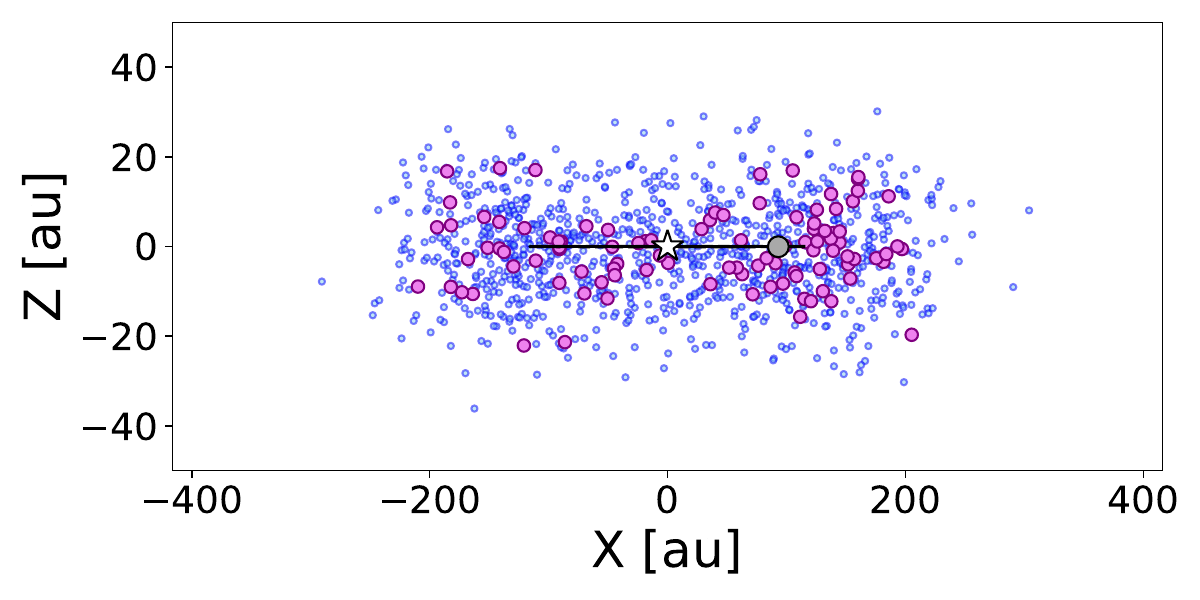}
    
    \caption{Snapshot at the final time step of the dynamical simulation from Figure \ref{fig:rms_e_i_M27_s4} (left), the derived surface brightness map following the method in Section \ref{sec:results_obs} (middle), and the simulated ALMA observations of the disc (right). The parameters for the synthetic observations are calculated using the Gaussian belt model and visibility fitting code presented in \citet{Marshall23}, and those results are presented in Tables \ref{tab:simobsc6-3} and \ref{tab:simobsc6-4}.}
    \label{fig:morphology}
\end{figure*}

We then simulate observations of the disc using the CASA {\em simobserve} task using parameters for the observation (e.g. array configuration, precipitable water vapour, hour angle) to emulate the conditions HD~16743 was observed in by \citet{Marshall23}, and scale the total integration time to match the noise level of the actual observations of HD~16743, to obtain the model image shown in the right panel of Figure \ref{fig:morphology}. In the left panel of the same Figure, it can be seen that the particles (TPs and DPs) have radially spread from their initial distributions (see Figure \ref{fig:fiducial_sim}) and now occupy orbits interior to the GP. One would naively expect a planet to rapidly clear its orbit, but in the model illustrated here, the GP has a relatively low mass ($M_{\rm GP} = 18~M_{\oplus}$) and is therefore inefficient at ejecting or clearing material from its orbit in just 80 Myr. Furthermore, the DPs within the belt are continually exciting additional TPs onto GP-crossing orbits \citep[e.g.,][]{Munoz18}. 

Those TPs inside the GP are dynamically excited, exhibiting a higher mean eccentricity and inclination than the bulk of the TPs in the disc. For some, their interaction with the GP, whose orbit has not changed much from its initial conditions, may be limited if they are protected by a resonant mechanism, allowing them to avoid strong interaction with the GP and thus maintain stable orbits \citep[e.g.,][]{Malhotra96}. Others, however, will experience active perturbations from the GP, forming an extended component similar to the scattered disc in our solar system \citep[e.g.,][]{Gomes08}. Although both these phenomena, familiar in the solar system, are present in our models, the available data for studying them in this work is limited. Therefore, we cannot explore their characteristics further.



The surface brightness maps were converted into fits images with world coordinate system information and then run through the CASA \textit{simobserve} task using the same array configuration (C6-3, beam FWHM $\simeq 1\farcs0$), comparable atmospheric conditions with the precipitable water vapour and hour angle taken from the original data, and the on-source integration time adjusted such that the noise of the output image matches the level in the original observations, i.e. 15~$\mu$Jy/beam. The visibilities were then extracted from the simulated observations and modelled using an inclined and rotated Gaussian belt following the method presented in \cite{Marshall23}, which we summarise here for completeness. In short, we fitted the real and imaginary components of the $uv$ visibilities using a Gaussian belt defined by its total flux density $f_{\rm disc}$, peak radius $\mu_{0}$, width (standard deviation) $\sigma$, disc inclination $\theta$, and position angle $\phi$. The maximum likelihood model estimation was made using {\sc emcee} to explore the parameter space with 10 walkers per free parameter and a total of 10,000 realisations to generate the posterior probability distributions. Values and uncertainties for each parameter were taken from the 16th, 50th, and 84th percentiles of the marginalised distributions. The results of this process for the grid of 36 simulations are presented in Table \ref{tab:simobsc6-3}.

There is a limit to the vertical resolution of the disc that can be achieved with the ALMA array configuration used in the original observations ($h \simeq 0.03$). This is apparent in several of the final models, with aspect ratios indistinguishable from the initial conditions. We therefore performed the same modelling exercise with the more extended C6-4 ALMA array configuration to produce higher resolution maps of the disc with beam FWHM $\simeq 0\farcs6$ but based on the same input models again being passed through the \textit{simobserve} task. We then infer the disc parameters through the same maximum likelihood estimation using {\sc emcee} to determine the change in dynamical state of the disc for these lower mass, broader planetesimal ring models over the duration of the dynamical simulations. The results of the higher resolution synthetic models are presented in Table \ref{tab:simobsc6-4}.

The radii ($\mu_{0}$), widths ($\sigma$), and orientations ($\theta$, $\phi$) of the model belts summarised in Tables \ref{tab:simobsc6-3} and \ref{tab:simobsc6-4} are all broadly consistent with the observed disc after 80~Myr integration. 

\begin{figure}
    \includegraphics[width=\columnwidth]{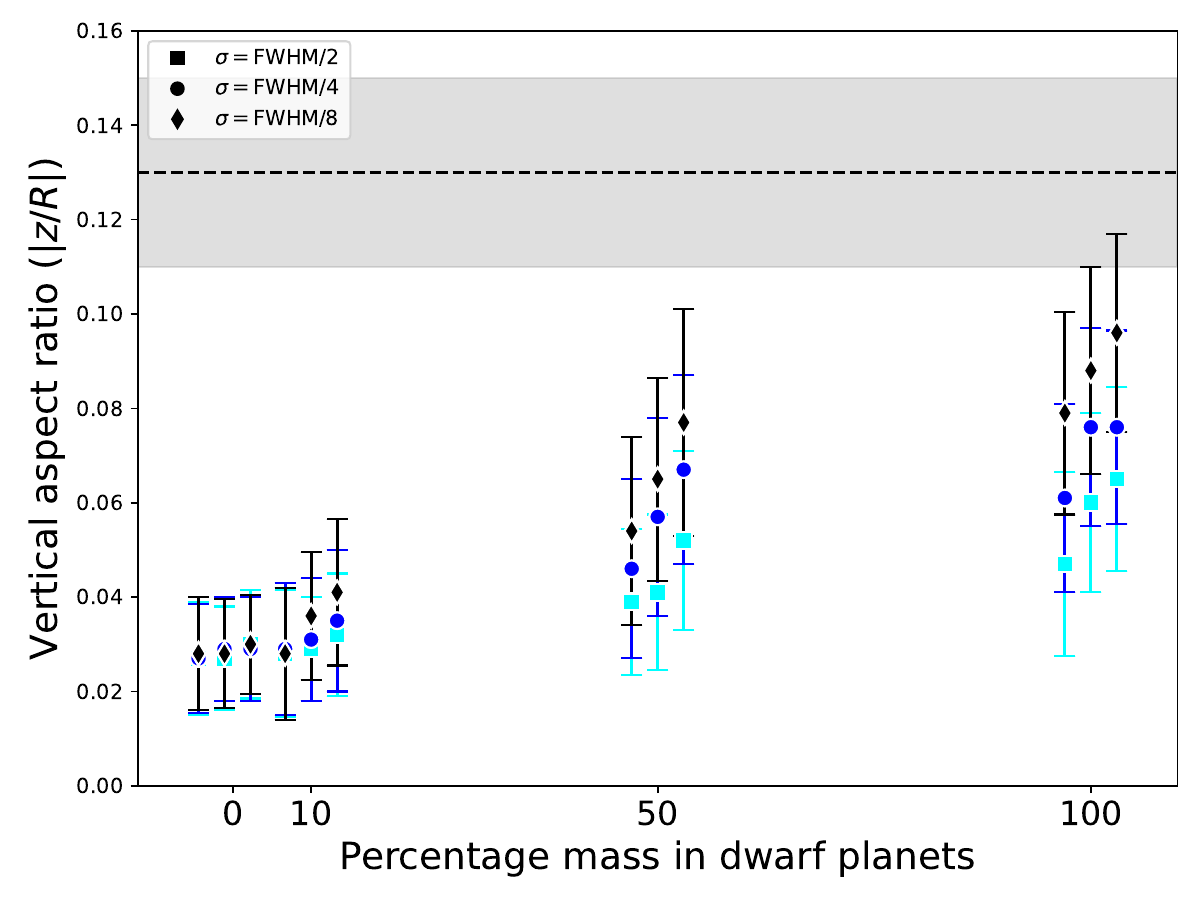}
    \caption{Vertical aspect ratio of the disc models from synthetic observations with the ALMA C6-4 configuration. The aspect ratios (and uncertainties, see Table \ref{tab:simobsc6-4}) are presented as a function of the percentage mass of the disc in dwarf planets. Simulations have been spread over $\pm$5\% along the $x$-axis for clarity. Data point colours denote the total mass of the system: 18 (cyan), 27 (blue), or 36 (black) $M_{\oplus}$. The horizontal dashed line and surrounding shaded region denote the observed vertical aspect ratio of HD~16743's disc and its associated uncertainty ($h = 0.13^{+0.02}_{-0.02}$). We see that the model aspect ratios approach values consistent with the observations for discs comprised solely of planetesimals with cumulative masses greater than or equal to 27~$M_{\oplus}$.}
    \label{fig:h_vs_mass}
\end{figure}

In Figure \ref{fig:h_vs_mass} we present the results of the simulated observations, visualising the end states of the HD~16743 model systems. We can see there is a clear trend in both overall disc mass and the fraction of disc mass in DPs, driving a larger vertical aspect ratio for the system. The belt width also has an impact (albeit weaker) on the final vertical aspect ratio of the synthetic disc. Model systems with initially narrower belt widths reach comparable vertical aspect ratios as more massive systems, or systems with higher mass fractions of DPs. The impact of the angular resolution on the determined parameters for each model is weak, with the results for each system being consistent with synthetic observations by either ALMA C6-4 or C6-3 array configurations. The uncertainties are slightly smaller at higher angular resolution, and the inferred vertical aspect ratios are slightly larger. Still, these differences are not significant and may be attributed to numerical precision. 

\section{Discussion}
\label{sec:discussion}

\subsection{Broader applicability}

We have considered the dynamical evolution of a set of model systems representing the outer belt of the HD~16743 debris disc. Our results demonstrate that, consistent with theoretical predictions, a self-stirring scenario where the belt architecture is driven by gravitational interactions between DPs and smaller bodies can explain the observed radial and vertical structure of the disc. Furthermore, we also found that the presence of a GP interior to the belt for the same total system mass can plausibly reproduce the observations, with the advantage of reducing the mass required within the debris disc by up to a factor of two. For HD~16743, the self-stirring mass has been estimated from 18 to 3,000~$M_{\oplus}$, depending on the stirring model invoked \citep{2012Pan,Pearce22}. Here we find that an intermediate mass of 36~$M_{\oplus}$ can broadly reproduce the disc structure within the lifetime of the system, intermediate to those values but orders of magnitude lower than the most conservative estimate. These findings may go some way to addressing the current mass problem of describing the architectures of debris discs through self-stirring \citep{WyattKrivov21}.

Another outcome of the analysis is that the DPs driving the dynamical excitation must (initially) be confined within a region much narrower than the observed belt width. After a rapid initial excitation (within 0.5 Myr, see Figure \ref{fig:rms_e_i_M27_s4}) the disc evolution follows the expected $t^{1/4}$ dependence on the increase in $e$ and $i$ for the DPs and TPs following \citet{Krivov18}, as shown in Figures \ref{fig:rms_e} and \ref{fig:rms_i}. However, toward the end of our integrations (80~Myr), we see a tail off in the excitation of the discs, particularly for those that are initially most massive and narrowest (see Figure \ref{fig:rms_final}). This indicates that the self-stirring scenario experiences diminishing returns and may eventually struggle to explain excited discs around older stars ($> 100~$Myr), as the DPs exciting these belts are spread and scattered, diluting their influence within the belt. Conversely, our findings suggest that mixed-stirring scenarios can aid in estimating the dynamical total masses of debris discs from their measured radial widths, particularly for narrow discs such as Fomalhaut and HR~4796 \citep{Macgregor17,Kennedy18}.

\subsection{Influence of modelling assumptions}

In constructing our model system to replicate the HD~16743 debris disc, we have necessarily made a number of simplifying assumptions regarding the dynamical state of the system and the forces acting on dust grains within the disc. 

We assume the disc is unstirred at the start of our integrations, with the DPs and TPs distributed within a tightly clustered range of $e$ and $i$ (Figure \ref{fig:fiducial_sim}). As can be seen in Figures \ref{fig:rms_e} and \ref{fig:rms_i}, the system undergoes a rapid period of excitement within the first (few) Myr after the simulations begin, before further change in the eccentricity and inclination plateau or taper off following an approximate $t^{1/4}$ trend. Instead, the DPs might be instantiated with some dynamically hotter distribution consistent with the vertical extent of protoplanetary discs \citep[e.g.][]{2020Villenave,2023Bae,2023Miotello} or the youngest debris discs \citep[$h \simeq 0.02$,][]{Terrill23}. In the absence of perturbing bodies, radiation pressure and collisions should produce a vertical aspect ratio of 0.04~$\pm$~0.02 at optical to mid-infrared wavelengths \citep{2009Thebault}. The appropriate initial conditions for the dynamical excitation of debris discs are still a matter of discussion. Beginning with a more evolved distribution (higher $e$, $i$), representative of real systems, would reduce the time required to excite the disc to the degree inferred from observations. However, starting with a dynamically cold system as we have here is a conservative assumption and places an upper limit on the time it would take to replicate the observed vertical stirring. Here we found that for this conservative assumption, we can still meet the observed conditions of HD~16743's disc within the time and mass constraints of the observations with our model grid.

Throughout the evolution of our models, we assume that the disc particles (DPs and TPs) do not undergo collisions. The largest (most massive) bodies in a collisional cascade hold the dominant fraction of the disc mass and are thereby responsible for the dynamical heating. These bodies are not directly involved in the collisional cascade, which only engages much smaller bodies, such that the most massive bodies will not undergo transformative (disruptive) collisions within the period of modelling simulated in this work, though scattering between DPs and TPs is essential to the evolution of the belt. Keeping both the total mass and the mass distribution of the DPs fixed within the simulations is thus a necessary approximation given collisions between DPs and TPs are being neglected, while the degree of stirring induced in the belt changes as the mass of DPs changes (see Section \ref{ssec:ndp}). Mass erosion would definitely reduce stirring, as well as the size reduction of the largest planetesimals caused by collisions. However, considering the time scale we are using and the mass density in the disc, we don't expect these effects to be significant. Furthermore, we do not consider non-gravitational forces acting upon DPs and TPs that may excite their orbits, such as radiation forces like the Yarkovsky effect on small planetesimals \citep{YORP}. 

We have omitted the effect of gas on the motion and settling of dust grains within the debris disc. As yet, there exists no detection of (CO) gas within HD~16743's debris disc \citep[predicted $M_{\rm CO} \approx 2.8 \times 10^{-7} M_{\oplus}$,][]{Kral2017}. The presence of a substantial gas disc would carry smaller micron-sized grains within it and cause larger mm-sized grains to decouple from the gas and settle into a vertically thinner disc\footnote{For HD~16743, the settling time (given the predicted CO gas mass) for 1~$\mu$m-sized grains is $\sim$1.7~Gyr, far older than the system.}. That the debris disc around HD~16743 has the same vertical extent at both near-infrared and millimetre wavelengths points to the absence of gas as a modifying factor in the dust motions \citep{Olofsson2022}.

\subsection{Dependence on the number of dwarf planets}
\label{ssec:ndp}
In a previous work, \cite{Munoz23} found a weak dependence of the disc stirring on the number of DPs. As the number of DPs increases, the level of stirring achieved certainly decreases. Specifically, increasing the number of DPs from 100 to 250 results in a reduction of stirring levels by a factor of 1.5. This suggests that the DPs in the simulations presented here can be viewed as pseudo-particles. Although it would be advantageous to include a larger number of DPs, computational constraints limit this possibility\footnote{Each dynamical simulation took $\simeq~24~$hrs to run for the 80~Myr integration time using a single core. Further post-processing to generate the ring models and simulated observations were a few hours per simulation.}. Given the modest decrease in stirring observed with a 2.5-fold increase in the number of particles, with the computational tools used in this paper, there is little motivation to conduct a study with a significantly greater number of DPs or a correspondingly longer integration time. Furthermore, in the Solar System, the largest 31 bodies in the Kuiper belt may account for a substantial fraction (5 to 90\%) of the total mass in the classical belt \citep[$M_{\rm KB} \simeq 0.01$--$0.20~M_{\oplus}$,][]{Bernstein04,Pitjeva18}, such that considering only their influence should be representative of the dominant bodies within the belt, beyond the giant planets. While the most common size of bodies within the Kuiper belt may be around 50-100 km, it is yet unclear if these bodies are large enough to significantly stir the belt. Additionally, extending the distribution of DPs to such levels is not computationally feasible, so the role of such smaller objects in interpreting the observations discussed here remains to be tested.

\subsection{Comparison with Krivov \& Booth 2018} 
\label{subsec:KB18_discussion}

The results for the growth of RMS$_{e}$ and RMS$_{i}$ presented in Figures \ref{fig:rms_e} and \ref{fig:rms_i} do not always follow the theoretical expectations derived from the calculations presented in \cite{Krivov18}. In particular, for the models with 100\% DPs stirring the disc, we would naively expect complete or very close agreement between the models and theory. In general, the degree of fidelity between our simulations and the theoretical predictions increases as the belt width decreases and the \% mass in DPs increases. However, the dynamical simulations presented here have several differences from those of \citet{Krivov18}. Firstly, the DPs in our simulations follow a power-law distribution in mass rather than being all the same mass. Secondly, the DPs have non-negligible, independent orbital eccentricities rather than the same assumed eccentricity regardless of mass. Thirdly, the stirring is taking place across a relatively broad range of radii ($\Delta R$ from 14 to 56 au) compared to the narrow annulus ($\Delta a = 10~$au) considered in \cite{Krivov18}.

As a point of comparison to the dynamical models presented here, we can use equations (5)--(10) from \cite{Krivov18} to calculate (i) the mass required to drive the disc aspect ratio to the observed value within 80~Myr and (ii) the time required for a disc of 18~$M_{\oplus}$ to drive the disc aspect ratio to the observed value. Consistent with the greatest stirring seen in our simulations, we assume a belt width  (their $\Delta a$) of $\mathrm{FWHM}/8$. In the first instance, we find that it would require a total disc mass of 72~$M_{\oplus}$ to reproduce the disc aspect ratio within 80~Myr, and for the second scenario, it would take an 18~$M_{\oplus}$ disc $\sim$320~Myr to excite the disc to the observed aspect ratio. This second result is interesting as the time required is still consistent with the (poorly constrained) upper limits to the age of the system, although far greater than the likely system age of 57~$\pm$~19~Myr \citep{Marshall23}. 

\subsection{Aspect ratio from simulations}

HD~16743's disc has an intermediate aspect ratio amongst the currently available values measured at millimetre wavelengths for debris discs \citep{Terrill23}. Its value is also consistent between near-infrared (1.6~$\mu$m) and millimetre (1.3~mm) wavelengths. Collisionally damped debris belts are expected to exhibit a wavelength dependence in their vertical aspect ratio \citep{2012Pan}, whereas belts with an aspect ratio independent of wavelength suggest collisional damping within the disc is inefficient \citep{2024Jankovic}. Thus, if the collisional velocities within the belt are low ($< 100~$m/s), but high enough to trigger a collisional cascade, the vertical aspect ratio could be independent of wavelength. However, our dynamical simulations show that DPs must be concentrated into a narrow annulus for the disc to acquire an aspect ratio consistent with observations, within the system's age, and with the inferred disc mass.

Two solutions to this issue present themselves: either the disc is already stirred at $t_{0}$ and has some non-zero initial aspect ratio (as discussed earlier), or the disc mass in DPs is indeed concentrated within a broader belt of dusty debris. In the first instance, the planetesimals within the disc are confined to a narrow region but with a non-zero excitation. This would be consistent with the invariance of $h$ with wavelength for this system, such that the observed aspect ratio is in fact primordial in origin. For the second case, the initially narrowly constrained planetesimals (DPs and TPs) spread, and the greater the degree of initial concentration, the greater the final belt width. This means that in order to match the radial width of the observations, some additional mechanism of confinement must be invoked, otherwise the resulting belt would be too broad compared to its observed radial extent.

A third uncertainty in the determination of the disc aspect ratio is its functional form. We have assumed a Gaussian vertical distribution for the dust in HD~16743's disc, consistent with previous works \citep[][ and references therein]{Terrill23}. However, in scattered light imaging of $\beta$ Pictoris' disc, the vertical aspect ratio is well matched by a combination of Lorentzian rather than Gaussian distributions \citep{Golimowski06,Lagrange12,2019Matra}. Higher angular resolution observations of HD~16743 may reveal that some other functional form is required to match the vertical distribution, leading to a change in the inferred aspect ratio and its interpretation \citep[see, e.g., the discussion in][]{Sefilian2025}.

\subsection{Constraints on planetary companions from the vertical aspect ratio}
\label{subsec:planet_alone_pred}

We have thus far examined the role of massive planetesimals within the HD~16743 disc in shaping its structure, both with and without an inner planetary companion. However, this analysis assumed a circular and coplanar planetary orbit. In principle, another mechanism that could explain -- or contribute to explaining -- the disc's thickness is its long-term, secular interactions with a \textit{misaligned} planet \citep[e.g.,][and references therein]{Wyatt99, Sefilian2025}. To explore this possibility, we assume a massless disc, i.e., neglecting self-stirring processes, and ask: what orbital parameters would allow an inner misaligned planet to reproduce the observed aspect ratio?

It is well-known that an initially misaligned planet--debris disc system evolves secularly toward alignment, provided the disc is massless \citep{Sefilian2025}. In this process, planetesimals throughout the disc undergo inclination oscillations between 0 (their initial inclination) and twice the initial mutual inclination $I_p(0)$ with the planet \citep{Murray99}. This occurs over secular timescales, $T_{\rm sec}(a_d) \propto a_d^{7/2}$, which increase with distance \citep[see equation (30) in][]{Sefilian2025}. After at least one secular timescale has passed at the disc’s outer edge, i.e., $T_{\rm{sec}}(a_{\rm out}) \leq t_{\rm age}$, the disc relaxes into a boxy, symmetric structure centered around the planetary orbit, with a distance-independent vertical aspect ratio $\mathcal{H}(R) \approx I_p(0)$ \citep{Sefilian2025}. Based on this, the measured aspect ratio of $\approx 0.13 \pm 0.02$ \citep{Marshall23} implies an initial misalignment of $\approx 7.45 \pm 1.15$ degrees. 
Furthermore, using the expression of $T_{\rm sec}(a_d)$ given in \citet[][]{Sefilian2025}, we find that an initially razor-thin disc would attain the observed aspect ratio for planetary masses $m_{\rm GP}$ exceeding:\footnote{For simplicity, Equation (\ref{eq:mp_ap_Md0_theory}) is presented assuming that the planetary semimajor axis is much smaller than the disc's outer edge ($a_{\rm GP}/ a_{\rm out} \lesssim 1$).}
\begin{equation}
    \frac{m_{\rm GP}}{M_J} \gtrsim 0.19 ~
    \bigg(\frac{100~{\rm Myr}}{t_{\rm age}} \bigg)
    \bigg(\frac{100~{\rm au}}{a_{\rm GP}}\bigg)^2  . 
    \label{eq:mp_ap_Md0_theory}
\end{equation}
In Equation~(\ref{eq:mp_ap_Md0_theory}), $a_{\rm GP}$ is the planetary semimajor axis, and we have used $M_c = 1.537 M_{\odot}$ and $a_{\rm out} = 197.4~{\rm au}$. 
Conversely, assuming a minimum system age of $12$ Myr \citep{Marshall23}, 
Equation~(\ref{eq:mp_ap_Md0_theory}) can be used to place an upper limit on $m_{\rm GP}$ as follows:
\begin{equation} 
{\rm max}[m_{\rm GP}(a_{\rm GP})] \approx 1.56 M_J ~ \bigg(\frac{100~{\rm au}}{a_{\rm GP}}\bigg)^2.
    \label{eq:map_mp_AAS}
\end{equation}
We note that Equations (\ref{eq:mp_ap_Md0_theory}) and (\ref{eq:map_mp_AAS}) are independent of $I_p(0)$\footnote{Strictly speaking, this statement holds only for small initial misalignments, as is the case here \cite[see, e.g.,][]{Laskar2010}.}, and we have assumed that the planet is on a circular orbit.

\begin{figure}
\includegraphics[width=\columnwidth]{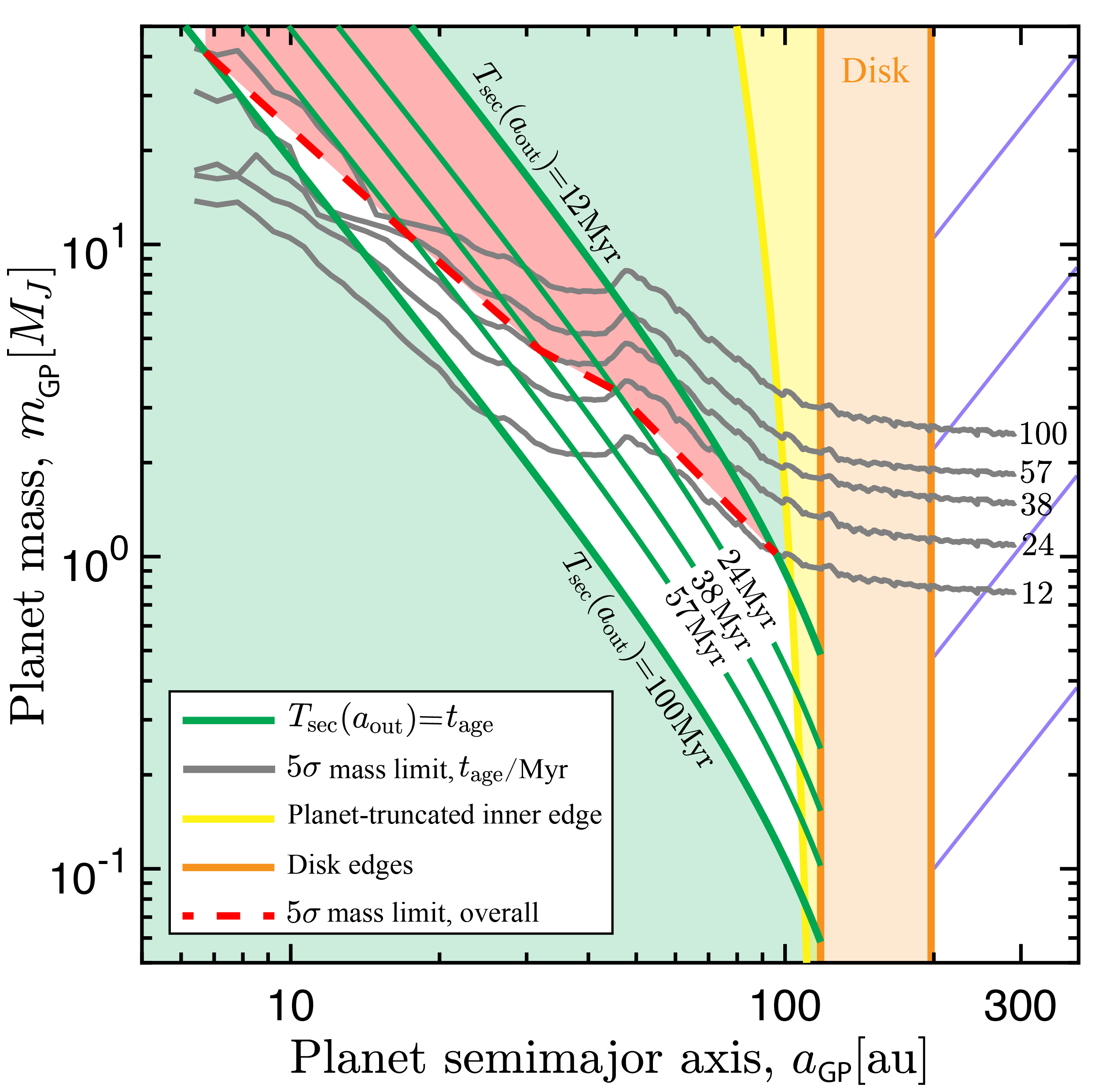}
\caption{The mass of a hypothetical planet $m_{\rm GP}$ as a function of its semimajor axis $a_{\rm GP}$ capable of vertically thickening an HD~16743-like debris disc within the system age. The possible combinations of $m_{\rm GP}$ and $a_{\rm GP}$ for a given system age $t_{\rm age}$ \citep[calculated using equation 30 from][]{Sefilian2025} are shown as green curves. 
Colour-shaded regions are excluded based on the following:
(i) the planet cannot reside within the disc (orange region); 
(ii) the yellow region represents where the disc's inner edge would have been carved at a larger distance than that observed; 
(iii) the green regions indicate secular timescales at the disc's outer edge that fall outside the system's age range of $12-100$ Myr; 
and
(iv) the gray curves show age-dependent upper limits on companion mass based on SPHERE observations \citep{Marshall23}.
The red region corresponds to the maximum dynamically inferred $m_{\rm GP}$ (Eq.~\ref{eq:mp_ap_Md0_theory}) that remains consistent with SPHERE observational limits.
The remaining white (unshaded) region indicates the range of parameters that satisfy all the aforementioned conditions, assuming the planet is interior to the disc (see the blue-hatched region).
See the text (Section \ref{subsec:planet_alone_pred}) for details.}
\label{fig:planet_mp_ap}
\end{figure}

Figure \ref{fig:planet_mp_ap} shows the required planet mass as a function of semimajor axis to thicken the disc vertically. Calculations are performed using equation (30) in \citet{Sefilian2025} for different values of $t_{\rm age}$, ranging from $12$ to $100$ Myr (green curves). This range is consistent with age estimates in the literature \citep{Marshall23}, although the upper limit of $100$ Myr is lower than the maximum value of $500$ Myr suggested by \citet{Desidera15}. 
Looking at Figure \ref{fig:planet_mp_ap}, it is evident that a misaligned companion can reproduce the observed aspect ratio within $\sim 12 - 100$ Myr, 
provided its semimajor axis lies between $\sim 7 $ and $110$ au
and its mass is between $25 M_{\earth}$ and $40 M_{J}$; this is highlighted as the white region in the figure. 
Note that for a given $t_{\rm age}$, the minimum $m_{\rm GP}$ is larger at smaller $a_{\rm GP}$ such that $m_{\rm GP} \propto a_{\rm GP}^{-2}$; see also Equation (\ref{eq:mp_ap_Md0_theory}).
This ``allowed'' range of planetary parameters is consistent with 
(i) current age-dependent observational limits on potential companions based on SPHERE data \citep[red-shaded area and gray curves;][]{Marshall23},
and 
(ii) dynamical arguments that the disc's inner edge is carved through overlapping first-order mean motion resonances. The latter is shown as a yellow curve in Figure \ref{fig:planet_mp_ap}, along which  $a_{\rm GP} + \Delta a_{\rm GP} = a_{\rm in}$ where $\Delta a_{\rm GP}$ is the width of the chaotic zone on either side of the planetary orbit \citep{Wisdom1980}:
\begin{equation}
    \Delta a_{\rm GP} \approx 1.3 a_{\rm GP}  \bigg(\frac{m_{\rm GP}}{M_c + m_{\rm GP}} \bigg)^{2/7} .
\end{equation}
Before moving on, we emphasize that the allowed planetary parameters are degenerate with the system’s uncertain age. Nevertheless, Figure \ref{fig:planet_mp_ap} clearly shows that the system cannot be younger than $12$ Myr -- assuming the disc is sculpted by this mechanism -- as a younger system would require a planet that should have already been detected.

In closing, we note that Figure \ref{fig:planet_mp_ap} is presented as an illustrative example of one possible planetary configuration and does not preclude other plausible scenarios. For instance, the system could host a companion exterior to the disc (i.e., within the blue-hatched region), with or without an additional inner companion. Therefore, the predictions discussed in this section should be interpreted with caution. The presence of exterior companions and/or multiple inner planets can significantly influence the disc dynamics \citep[][]{Nesvold2017, Dong+2020, Farhat23}, let alone the long-term gravitational role of the disc itself -- an effect that has been shown to be important even when the disc is less massive than the perturbing companion \citep{SefilianTouma2019, Sefilian21, Sefilian2023, Sefilian2025, Poblete+2023}. Expanding any further on these points is beyond the scope of the present study. 


\section{Conclusions}
\label{sec:conclusions}

In this work, we investigated the impact of a combination of gravitational perturbations on the evolution of an initially cold debris disc that is formed by 1,000 test particles. The perturbations include a giant planet located interior to the disc and 100 dwarf planets (DPs) distributed throughout the disc itself. We examined three values of dynamical mass (18, 27, and 36 $M_\oplus$) spanning the range required to trigger disc stirring, according to the model of \citet{2012Pan}. 

We explored various configurations, ranging from scenarios where all the mass is concentrated in the giant planet, meaning the DPs do not influence the disc's evolution, to cases where all the mass is held by the DPs, resulting in a situation without a giant planet. This latter scenario corresponds to the well-known self-stirring model. In addition, we analysed two intermediate models: one where 90\% of the mass is in the giant planet and 10\% in the DPs, and another where 50\% of the mass is in the giant planet and 50\% in the DPs. These intermediate configurations represent different setups of the ``mixed-stirring'' model discussed by \citet{Munoz15,Munoz23}.

We found that models considering only the giant planet are inefficient at stirring the full planetesimal belt and insufficient to reproduce the observed disc stirring to the level observed in the ALMA images. By contrast, self-stirring equivalent models show the greatest increase in inclination, such that the disc may be presumed to be self-stirring. However, our mixed-stirring scenarios are almost as effective as the purely self-stirring model, especially the 50\%-50\% models. 

From the models considered here, we find that 50\% or 100\% DPs simulations with $M_{\rm tot}~\geq~27~M_{\oplus}$ and $\sigma~\leq~\mathrm{FWHM}/4$ are able to reproduce the observed architecture within the assumed 80~Myr timescale (i.e. they are consistent within 3-$\sigma$). The most vertically extended models are also the most radially extended, such that the addition of greater mass to the system to reproduce exactly the vertical extent of the disc may result in a final architecture that is too broad compared to the observations. The inclination only gradually increases with age after an initial period of dynamical heating, such that increasing the integration time would not resolve the mismatch between modelled and observed aspect ratios, particularly for systems with tight upper bounds on their ages (e.g., debris discs around stellar moving group members).

Mixed-stirring can therefore produce an equivalent stirring with just half the disc mass required compared to self-stirring, addressing a current perceived problem in the mass budget required for debris discs \citep{WyattKrivov21}.

\section*{Acknowledgements}

The authors thank the referee, Dr. Virginie Faramaz-Gorka, for their helpful comments on the paper. This research has made use of the SIMBAD database, operated at CDS, Strasbourg, France \citep{2000Wenger}. This research has made use of the Astrophysics Data System, funded by NASA under Cooperative Agreement 80NSSC21M00561.
JPM acknowledges research support by the National Science and Technology Council of Taiwan under grant NSTC 112-2112-M-001-032-MY3.
AAS is supported by the Heising-Simons Foundation through a 51 Pegasi b Fellowship.

\textit{Facilities:} None.

\textit{Software:} Simulations in this paper made use of the $N$-body code {\sc rebound} \citep{rebound}. The simulations were integrated using the hybrid symplectic integrator {\sc mercurius} \citep{reboundmercurius}. The SimulationArchive format was used to store fully reproducible simulation data \citep{reboundsa}. This paper has made use of the Python packages {\sc astropy} \citep{AstroPy13,AstroPy18,AstroPy22}, {\sc scipy} \citep{Virtanen20}, {\sc numpy} \citep{Harris20}, {\sc matplotlib} \citep{Hunter07}, {\sc emcee} \citep{ForemanMackey13}, {\sc corner} \citep{ForemanMackey16}, and {\sc galario} \citep{Tazzari18}.

\section*{Data Availability}

The data underlying this article are available in the article and its online supplementary material. The supplementary data and analysis scripts are provided for public access on \href{https://figshare.com/projects/HD_16743_Dynamics_Paper/180316}{Figshare}.



\bibliographystyle{mnras}
\bibliography{hd16743} 


\appendix

\section{Simulated ALMA observations}

In this Appendix, we present the tabulated results of modelling the simulated observations of our dynamical simulations following the process summarised in Section \ref{sec:results_obs}. The fitting for the ALMA C6-3 array configuration is presented in Table \ref{tab:simobsc6-3}, whilst the fitting for the C6-4 array configuration is presented in Table \ref{tab:simobsc6-4}.

\begin{table*}
    \centering
    \caption{Comparison of simulated ALMA observations of dynamical models using the C6-3 array configuration ($\theta_{\rm beam} \simeq 1\arcsec$) with the original observations from \citet{Marshall23}. The (marginalised) posteriors and chains for each simulated observation are available on the Figshare repository. \label{tab:simobsc6-3}}
    \begin{tabular}{ccccccccc}
    \hline
    DPs & Mass & Width & $t_{\rm sim}$ & $\mu_{0}$ & $\sigma$ & $h$ & $\theta$ & $\phi$  \\
    (\%)    & ($M_{\oplus}$)   & (FWHM)  & (Myr) & (au)   &  (au)    & & ($\degr$) & ($\degr$) \\
    \hline
    \multicolumn{4}{c}{Observations} & 157.7$^{+2.6}_{-1.5}$ & 33.7$^{+3.4}_{-3.3}$ & 0.13$^{+0.01}_{-0.02}$ & 87.3$^{+1.9}_{-2.5}$ & 168.5$^{+0.6}_{-0.5}$ \\
    \hline 
    0 & 18 & 1/2 & 0 & 145.9$^{+2.1}_{-2.1}$ &  29.5$^{+2.9}_{-2.5}$ &  0.03$^{+0.01}_{-0.01}$ &  88.2$^{+1.1}_{-1.1}$ & 168.5$^{+0.4}_{-0.4}$\\
    0 & 18 & 1/4 & 0 & 151.2$^{+1.6}_{-1.4}$ &  17.3$^{+2.7}_{-2.8}$ &  0.03$^{+0.01}_{-0.01}$ &  88.3$^{+1.1}_{-1.1}$ & 168.5$^{+0.3}_{-0.3}$\\
    0 & 18 & 1/8 & 0 & 153.4$^{+1.3}_{-1.3}$ &   8.1$^{+3.4}_{-3.2}$ &  0.03$^{+0.01}_{-0.01}$ &  88.3$^{+1.0}_{-1.2}$ & 168.5$^{+0.3}_{-0.3}$\\
    \hline
    0 & 18 & 1/2 & 80 & 146.2$^{+2.2}_{-2.1}$ &  30.3$^{+2.9}_{-2.6}$ &  0.03$^{+0.01}_{-0.01}$ &  88.1$^{+1.1}_{-1.1}$ & 168.5$^{+0.4}_{-0.4}$\\
    0 & 18 & 1/4 & 80 & 151.5$^{+1.5}_{-1.6}$ &  17.2$^{+2.9}_{-2.9}$ &  0.03$^{+0.01}_{-0.01}$ &  88.3$^{+1.1}_{-1.2}$ & 168.4$^{+0.3}_{-0.3}$\\
    0 & 18 & 1/8 & 80 & 153.6$^{+1.4}_{-1.3}$ &   7.6$^{+3.2}_{-2.9}$ &  0.03$^{+0.01}_{-0.01}$ &  88.4$^{+0.9}_{-1.1}$ & 168.5$^{+0.3}_{-0.3}$\\
    10 & 18 & 1/2 & 80 & 147.1$^{+2.0}_{-2.1}$ &  29.8$^{+2.7}_{-2.5}$ &  0.03$^{+0.01}_{-0.01}$ &  88.2$^{+1.1}_{-1.3}$ & 168.5$^{+0.4}_{-0.3}$\\
    10 & 18 & 1/4 & 80 & 157.2$^{+1.6}_{-1.6}$ &  18.8$^{+2.5}_{-2.8}$ &  0.03$^{+0.01}_{-0.01}$ &  87.9$^{+1.3}_{-1.2}$ & 168.5$^{+0.3}_{-0.3}$\\
    10 & 18 & 1/8 & 80 & 160.8$^{+1.7}_{-1.6}$ &  20.6$^{+2.9}_{-2.7}$ &  0.03$^{+0.01}_{-0.01}$ &  88.2$^{+1.2}_{-1.3}$ & 168.5$^{+0.4}_{-0.3}$\\
    50 & 18 & 1/2 & 80 & 146.3$^{+2.6}_{-2.7}$ &  37.2$^{+3.1}_{-2.9}$ &  0.04$^{+0.02}_{-0.02}$ &  87.5$^{+1.6}_{-1.5}$ & 168.5$^{+0.4}_{-0.4}$\\
    50 & 18 & 1/4 & 80 & 152.1$^{+2.1}_{-2.4}$ &  29.3$^{+2.6}_{-2.9}$ &  0.04$^{+0.02}_{-0.02}$ &  87.3$^{+1.6}_{-1.4}$ & 168.5$^{+0.4}_{-0.4}$\\
    50 & 18 & 1/8 & 80 & 152.5$^{+2.7}_{-2.9}$ &  34.0$^{+2.9}_{-3.4}$ &  0.05$^{+0.02}_{-0.02}$ &  86.9$^{+1.8}_{-1.5}$ & 168.6$^{+0.5}_{-0.4}$\\
    100 & 18 & 1/2 & 80 & 145.5$^{+3.0}_{-2.7}$ &  40.0$^{+3.3}_{-3.2}$ &  0.05$^{+0.02}_{-0.02}$ &  86.7$^{+2.0}_{-1.3}$ & 168.5$^{+0.4}_{-0.4}$\\
    100 & 18 & 1/4 & 80 & 151.4$^{+2.6}_{-2.6}$ &  34.9$^{+3.4}_{-3.2}$ &  0.06$^{+0.02}_{-0.02}$ &  86.7$^{+1.9}_{-1.5}$ & 168.5$^{+0.4}_{-0.4}$\\
    100 & 18 & 1/8 & 80 & 151.8$^{+2.4}_{-2.6}$ &  32.2$^{+2.8}_{-3.1}$ &  0.06$^{+0.02}_{-0.03}$ &  86.7$^{+2.1}_{-1.6}$ & 168.5$^{+0.4}_{-0.4}$\\
    \hline
    10 & 27 & 1/2 & 80 & 148.9$^{+2.1}_{-2.1}$ &  30.4$^{+2.7}_{-2.6}$ &  0.03$^{+0.02}_{-0.01}$ &  88.1$^{+1.2}_{-1.3}$ & 168.5$^{+0.4}_{-0.4}$\\
    10 & 27 & 1/4 & 80 & 158.4$^{+1.6}_{-1.7}$ &  20.2$^{+2.7}_{-2.7}$ &  0.03$^{+0.01}_{-0.01}$ &  87.8$^{+1.4}_{-1.2}$ & 168.5$^{+0.3}_{-0.3}$\\
    10 & 27 & 1/8 & 80 & 162.8$^{+1.9}_{-2.1}$ &  25.4$^{+3.2}_{-3.1}$ &  0.04$^{+0.01}_{-0.02}$ &  87.6$^{+1.6}_{-1.3}$ & 168.5$^{+0.3}_{-0.4}$\\
    50 & 27 & 1/2 & 80 & 147.6$^{+2.8}_{-2.9}$ &  38.6$^{+3.3}_{-3.2}$ &  0.05$^{+0.02}_{-0.02}$ &  87.1$^{+1.8}_{-1.5}$ & 168.5$^{+0.4}_{-0.4}$\\
    50 & 27 & 1/4 & 80 & 153.7$^{+2.9}_{-2.8}$ &  39.1$^{+2.9}_{-3.3}$ &  0.05$^{+0.02}_{-0.02}$ &  86.5$^{+1.9}_{-1.4}$ & 168.5$^{+0.5}_{-0.4}$\\
    50 & 27 & 1/8 & 80 & 151.5$^{+2.9}_{-2.9}$ &  41.4$^{+3.5}_{-3.3}$ &  0.06$^{+0.02}_{-0.02}$ &  86.2$^{+2.1}_{-1.5}$ & 168.5$^{+0.5}_{-0.5}$\\
    100 & 27 & 1/2 & 80 & 145.3$^{+3.3}_{-3.1}$ &  44.2$^{+3.5}_{-3.8}$ &  0.06$^{+0.02}_{-0.02}$ &  86.3$^{+2.1}_{-1.6}$ & 168.5$^{+0.5}_{-0.5}$\\
    100 & 27 & 1/4 & 80 & 151.7$^{+3.1}_{-3.1}$ &  39.9$^{+3.2}_{-3.4}$ &  0.07$^{+0.02}_{-0.03}$ &  85.7$^{+2.3}_{-1.6}$ & 168.5$^{+0.5}_{-0.5}$\\
    100 & 27 & 1/8 & 80 & 152.3$^{+2.9}_{-3.0}$ &  36.1$^{+3.1}_{-3.1}$ &  0.08$^{+0.02}_{-0.03}$ &  86.3$^{+2.2}_{-1.9}$ & 168.5$^{+0.5}_{-0.5}$\\
    \hline
    10 & 36 & 1/2 & 80 & 147.7$^{+2.3}_{-2.4}$ &  31.6$^{+2.8}_{-2.9}$ &  0.03$^{+0.02}_{-0.01}$ &  87.8$^{+1.3}_{-1.2}$ & 168.5$^{+0.4}_{-0.4}$\\
    10 & 36 & 1/4 & 80 & 160.3$^{+1.9}_{-1.9}$ &  23.7$^{+2.8}_{-2.7}$ &  0.03$^{+0.02}_{-0.01}$ &  87.8$^{+1.3}_{-1.3}$ & 168.5$^{+0.3}_{-0.4}$\\
    10 & 36 & 1/8 & 80 & 164.8$^{+2.2}_{-2.6}$ &  28.5$^{+3.3}_{-3.0}$ &  0.04$^{+0.01}_{-0.02}$ &  87.3$^{+1.6}_{-1.2}$ & 168.5$^{+0.4}_{-0.4}$\\
    50 & 36 & 1/2 & 80 & 147.4$^{+3.2}_{-3.1}$ &  43.2$^{+3.6}_{-3.7}$ &  0.05$^{+0.02}_{-0.02}$ &  86.7$^{+1.9}_{-1.5}$ & 168.5$^{+0.5}_{-0.5}$\\
    50 & 36 & 1/4 & 80 & 152.9$^{+3.2}_{-3.3}$ &  44.0$^{+3.7}_{-3.9}$ &  0.07$^{+0.02}_{-0.02}$ &  86.7$^{+2.0}_{-1.6}$ & 168.5$^{+0.5}_{-0.5}$\\
    50 & 36 & 1/8 & 80 & 147.7$^{+3.5}_{-3.2}$ &  45.2$^{+3.9}_{-3.9}$ &  0.07$^{+0.02}_{-0.03}$ &  86.2$^{+2.4}_{-1.8}$ & 168.5$^{+0.5}_{-0.5}$\\
    100 & 36 & 1/2 & 80 & 142.8$^{+3.4}_{-3.4}$ &  46.2$^{+4.1}_{-3.8}$ &  0.08$^{+0.02}_{-0.03}$ &  86.0$^{+2.3}_{-1.8}$ & 168.4$^{+0.6}_{-0.5}$\\
    100 & 36 & 1/4 & 80 & 150.5$^{+3.4}_{-3.4}$ &  43.5$^{+3.9}_{-3.4}$ &  0.08$^{+0.02}_{-0.03}$ &  85.3$^{+2.6}_{-1.8}$ & 168.5$^{+0.5}_{-0.6}$\\
    100 & 36 & 1/8 & 80 & 154.0$^{+3.4}_{-3.3}$ &  40.3$^{+3.5}_{-3.6}$ &  0.10$^{+0.02}_{-0.02}$ &  86.0$^{+2.4}_{-2.1}$ & 168.4$^{+0.6}_{-0.5}$\\
    \hline
    \end{tabular}
\end{table*}

\begin{table*}
    \centering
    \caption{Similar to Table \ref{tab:simobsc6-3}, but using the C6-4 array configuration ($\theta_{\rm beam} \simeq 0\farcs6$). The (marginalised) posteriors and chains for each simulated observation are available on the Figshare repository. \label{tab:simobsc6-4}}
    \begin{tabular}{ccccccccc}
    \hline
    DPs & Mass & Width & $t_{\rm sim}$ & $\mu_{0}$ & $\sigma$ & $h$ & $\theta$ & $\phi$  \\
    (\%)    & ($M_{\oplus}$)   & (FWHM)  & (Myr) & (au)   &  (au)    & & ($\degr$) & ($\degr$) \\
    \hline
    \multicolumn{4}{c}{Observations} & 157.7$^{+2.6}_{-1.5}$ & 33.7$^{+3.4}_{-3.3}$ & 0.13$^{+0.01}_{-0.02}$ & 87.3$^{+1.9}_{-2.5}$ & 168.5$^{+0.6}_{-0.5}$ \\
    \hline
    0 & 18 & 1/2 & 0 & 146.2$^{+2.7}_{-2.6}$ & 30.0$^{+3.2}_{-3.4}$ & 0.03$^{+0.01}_{-0.01}$ & 88.3$^{+1.1}_{-1.1}$ & 168.5$^{+0.4}_{-0.4}$\\
    0 & 18 & 1/4 & 0 & 151.5$^{+1.8}_{-1.8}$ & 17.4$^{+3.0}_{-3.0}$ & 0.03$^{+0.01}_{-0.01}$ & 88.3$^{+1.0}_{-1.1}$ & 168.4$^{+0.3}_{-0.3}$\\
    0 & 18 & 1/8 & 0 & 153.2$^{+1.6}_{-1.5}$ &  7.6$^{+3.5}_{-3.0}$ & 0.03$^{+0.01}_{-0.01}$ & 88.3$^{+1.1}_{-1.1}$ & 168.5$^{+0.3}_{-0.3}$\\
    \hline
    0 & 18 & 1/2 & 80 & 145.9$^{+2.6}_{-2.5}$ & 30.0$^{+3.6}_{-3.4}$ & 0.03$^{+0.01}_{-0.01}$ & 88.2$^{+1.1}_{-1.1}$ & 168.5$^{+0.4}_{-0.4}$\\
    0 & 18 & 1/4 & 80 & 151.5$^{+1.9}_{-1.9}$ & 16.9$^{+3.0}_{-3.0}$ & 0.03$^{+0.01}_{-0.01}$ & 88.4$^{+1.0}_{-1.1}$ & 168.5$^{+0.3}_{-0.3}$\\
    0 & 18 & 1/8 & 80 & 153.6$^{+1.6}_{-1.6}$ &  8.3$^{+3.3}_{-3.3}$ & 0.03$^{+0.01}_{-0.01}$ & 88.4$^{+1.0}_{-1.1}$ & 168.5$^{+0.3}_{-0.3}$\\
    10 & 18 & 1/2 & 80 & 147.1$^{+3.0}_{-2.7}$ & 29.9$^{+3.2}_{-3.3}$ & 0.03$^{+0.01}_{-0.01}$ & 88.1$^{+1.1}_{-1.2}$ & 168.5$^{+0.4}_{-0.4}$\\
    10 & 18 & 1/4 & 80 & 157.2$^{+2.0}_{-2.1}$ & 18.5$^{+3.2}_{-3.3}$ & 0.03$^{+0.01}_{-0.01}$ & 88.1$^{+1.1}_{-1.2}$ & 168.5$^{+0.3}_{-0.3}$\\
    10 & 18 & 1/8 & 80 & 161.1$^{+2.2}_{-2.1}$ & 20.5$^{+3.4}_{-3.4}$ & 0.03$^{+0.01}_{-0.01}$ & 88.0$^{+1.2}_{-1.3}$ & 168.5$^{+0.4}_{-0.3}$\\
    50 & 18 & 1/2 & 80 & 146.0$^{+3.6}_{-3.5}$ & 37.5$^{+3.9}_{-3.9}$ & 0.04$^{+0.02}_{-0.02}$ & 87.5$^{+1.5}_{-1.3}$ & 168.5$^{+0.5}_{-0.5}$\\
    50 & 18 & 1/4 & 80 & 152.1$^{+2.7}_{-2.8}$ & 29.0$^{+3.3}_{-3.5}$ & 0.04$^{+0.01}_{-0.02}$ & 87.5$^{+1.5}_{-1.4}$ & 168.5$^{+0.4}_{-0.4}$\\
    50 & 18 & 1/8 & 80 & 152.6$^{+3.4}_{-3.3}$ & 34.5$^{+3.6}_{-3.2}$ & 0.05$^{+0.02}_{-0.02}$ & 87.1$^{+1.9}_{-1.6}$ & 168.5$^{+0.5}_{-0.5}$\\
    100 & 18 & 1/2 & 80 & 145.4$^{+3.6}_{-3.9}$ & 40.1$^{+4.0}_{-4.0}$ & 0.05$^{+0.02}_{-0.02}$ & 86.7$^{+1.8}_{-1.3}$ & 168.5$^{+0.5}_{-0.5}$\\
    100 & 18 & 1/4 & 80 & 150.9$^{+3.7}_{-3.6}$ & 35.1$^{+4.1}_{-3.7}$ & 0.06$^{+0.02}_{-0.02}$ & 86.6$^{+2.0}_{-1.4}$ & 168.5$^{+0.5}_{-0.6}$\\
    100 & 18 & 1/8 & 80 & 151.4$^{+3.1}_{-3.3}$ & 32.4$^{+3.8}_{-3.8}$ & 0.07$^{+0.02}_{-0.02}$ & 86.6$^{+2.0}_{-1.6}$ & 168.4$^{+0.5}_{-0.5}$\\
    \hline
    10 & 27 & 1/2 & 80 & 148.4$^{+2.6}_{-2.7}$ & 30.2$^{+3.5}_{-3.4}$ & 0.03$^{+0.02}_{-0.01}$ & 88.0$^{+1.2}_{-1.2}$ & 168.4$^{+0.5}_{-0.4}$\\
    10 & 27 & 1/4 & 80 & 158.2$^{+2.2}_{-2.1}$ & 19.9$^{+3.1}_{-2.8}$ & 0.03$^{+0.01}_{-0.01}$ & 87.8$^{+1.2}_{-1.1}$ & 168.5$^{+0.4}_{-0.4}$\\
    10 & 27 & 1/8 & 80 & 162.9$^{+2.6}_{-2.4}$ & 25.2$^{+3.7}_{-3.7}$ & 0.04$^{+0.02}_{-0.01}$ & 87.7$^{+1.3}_{-1.3}$ & 168.5$^{+0.4}_{-0.4}$\\
    50 & 27 & 1/2 & 80 & 147.8$^{+3.7}_{-3.8}$ & 39.1$^{+4.1}_{-4.0}$ & 0.05$^{+0.02}_{-0.02}$ & 87.1$^{+1.8}_{-1.4}$ & 168.5$^{+0.5}_{-0.5}$\\
    50 & 27 & 1/4 & 80 & 154.1$^{+3.7}_{-3.7}$ & 39.0$^{+4.0}_{-3.9}$ & 0.06$^{+0.02}_{-0.03}$ & 86.7$^{+1.9}_{-1.4}$ & 168.5$^{+0.5}_{-0.5}$\\
    50 & 27 & 1/8 & 80 & 151.6$^{+4.7}_{-4.1}$ & 41.6$^{+4.5}_{-4.0}$ & 0.07$^{+0.02}_{-0.02}$ & 86.5$^{+2.2}_{-1.8}$ & 168.5$^{+0.6}_{-0.6}$\\
    100 & 27 & 1/2 & 80 & 145.1$^{+4.2}_{-3.9}$ & 44.0$^{+4.8}_{-4.4}$ & 0.06$^{+0.02}_{-0.02}$ & 86.1$^{+2.2}_{-1.5}$ & 168.6$^{+0.6}_{-0.6}$\\
    100 & 27 & 1/4 & 80 & 151.8$^{+4.1}_{-4.1}$ & 40.6$^{+4.3}_{-4.2}$ & 0.08$^{+0.02}_{-0.02}$ & 85.9$^{+2.3}_{-1.7}$ & 168.5$^{+0.6}_{-0.5}$\\
    100 & 27 & 1/8 & 80 & 151.7$^{+3.9}_{-3.8}$ & 36.7$^{+4.3}_{-3.8}$ & 0.08$^{+0.02}_{-0.02}$ & 86.1$^{+2.3}_{-1.7}$ & 168.6$^{+0.6}_{-0.6}$\\
    \hline
    10 & 36 & 1/2 & 80 & 147.3$^{+2.7}_{-2.8}$ & 31.3$^{+3.5}_{-3.5}$ & 0.03$^{+0.02}_{-0.01}$ & 87.8$^{+1.2}_{-1.1}$ & 168.5$^{+0.4}_{-0.4}$\\
    10 & 36 & 1/4 & 80 & 160.2$^{+2.4}_{-2.3}$ & 23.8$^{+3.2}_{-3.2}$ & 0.04$^{+0.01}_{-0.01}$ & 87.8$^{+1.4}_{-1.3}$ & 168.5$^{+0.4}_{-0.4}$\\
    10 & 36 & 1/8 & 80 & 164.3$^{+2.8}_{-3.1}$ & 28.0$^{+3.9}_{-3.8}$ & 0.04$^{+0.01}_{-0.02}$ & 87.3$^{+1.5}_{-1.3}$ & 168.5$^{+0.4}_{-0.4}$\\
    50 & 36 & 1/2 & 80 & 147.2$^{+4.0}_{-3.9}$ & 43.3$^{+4.7}_{-4.3}$ & 0.05$^{+0.02}_{-0.02}$ & 86.6$^{+2.0}_{-1.4}$ & 168.5$^{+0.5}_{-0.5}$\\
    50 & 36 & 1/4 & 80 & 152.7$^{+4.3}_{-4.4}$ & 44.3$^{+4.6}_{-4.2}$ & 0.07$^{+0.02}_{-0.03}$ & 86.4$^{+2.2}_{-1.7}$ & 168.5$^{+0.5}_{-0.6}$\\
    50 & 36 & 1/8 & 80 & 147.2$^{+4.6}_{-4.6}$ & 45.1$^{+4.7}_{-4.4}$ & 0.08$^{+0.02}_{-0.03}$ & 85.9$^{+2.6}_{-1.8}$ & 168.5$^{+0.6}_{-0.6}$\\
    100 & 36 & 1/2 & 80 & 142.3$^{+4.6}_{-4.2}$ & 46.6$^{+4.9}_{-4.7}$ & 0.08$^{+0.02}_{-0.03}$ & 85.9$^{+2.3}_{-1.9}$ & 168.6$^{+0.6}_{-0.6}$\\
    100 & 36 & 1/4 & 80 & 149.9$^{+4.5}_{-4.4}$ & 43.3$^{+4.9}_{-4.3}$ & 0.09$^{+0.02}_{-0.03}$ & 85.8$^{+2.5}_{-2.0}$ & 168.5$^{+0.6}_{-0.7}$\\
    100 & 36 & 1/8 & 80 & 152.9$^{+4.6}_{-4.5}$ & 41.1$^{+4.8}_{-4.7}$ & 0.10$^{+0.02}_{-0.02}$ & 85.7$^{+2.7}_{-2.2}$ & 168.5$^{+0.6}_{-0.6}$\\
    \hline
    \end{tabular}
\end{table*}

\bsp	
\label{lastpage}
\end{document}